\documentclass[aps,prl,twocolumn,nofootinbib,groupedaddress,superscriptaddress
,longbibliography]{revtex4-2}
\AtBeginDocument{
\heavyrulewidth=.08em
\lightrulewidth=.05em
\cmidrulewidth=.03em
\belowrulesep=.65ex
\belowbottomsep=0pt
\aboverulesep=.4ex
\abovetopsep=0pt
\cmidrulesep=\doublerulesep
\cmidrulekern=.5em
\defaultaddspace=.5em
}

\usepackage{float}
\usepackage{comment}
\usepackage{booktabs}
\usepackage[percent]{overpic}
\usepackage{microtype}
\usepackage{array}
\usepackage{physics}
\usepackage{graphicx}
\usepackage{bbold}
\usepackage[usenames,dvipsnames]{xcolor}
\usepackage{tikz}
\usepackage{epsfig}
\usetikzlibrary{arrows.meta, automata, positioning, quotes, shapes}

\definecolor{ForestGreen}{HTML}{668000}
\definecolor{red1}{HTML}{FF4136}
\definecolor{green1}{HTML}{00802b}

\usepackage{amsmath,amssymb}
\usepackage{multirow}
\usepackage{bm}
\usepackage{mathtools}
\usepackage{dsfont}
\usepackage{amsfonts}
\usepackage[normalem]{ulem}
\usepackage{url}
\usepackage{wasysym}
\usepackage{multibib}
\newcites{SM}{Supplementary References} 
\usepackage[colorlinks=true,linkcolor=magenta,citecolor=magenta,hypertexnames=false]{hyperref}
\def\ba#1\ea{\begin{align}#1\end{align}}
\def\bg#1\eg{\begin{gather}#1\end{gather}}
\def\bpm{\begin{pmatrix}}
\def\epm{\end{pmatrix}}

\newcommand{\magenta}[1]{\textcolor{magenta}{#1}}

\newcommand{\ourtitle}{Emergent quantum Majorana metal from a chiral spin liquid}
\allowdisplaybreaks

\begin{document}
\title{\textbf{\ourtitle}}
%%%%%%%%%%%%

%%%%%%%%%%%%
\author{Penghao Zhu}
\altaffiliation{These authors contributed equally to this work.}
\affiliation{Department of Physics, The Ohio State University, Columbus, Ohio 43210, USA}
\author{Shi Feng}
\altaffiliation{These authors contributed equally to this work.}
\affiliation{Department of Physics, The Ohio State University, Columbus, Ohio 43210, USA}
\affiliation{Technical University of Munich, TUM School of Natural Sciences, Physics Department, 85748 Garching, Germany}
\affiliation{Munich Center for Quantum Science and Technology (MCQST), Schellingstr. 4, 80799 M{\"u}nchen, Germany}

\author{Kang Wang}
\altaffiliation{These authors contributed equally to this work.}
\affiliation{Beijing National Laboratory for Condensed Matter Physics and Institute of Physics,
Chinese Academy of Sciences, Beijing 100190, China}
\affiliation{School of Physical Sciences, University of Chinese Academy of Sciences, Beijing 100049, China}
\author{Tao Xiang}
\thanks{To whom correspondence should be addressed: txiang@iphy.ac.cn, trivedi.15@osu.edu}
\affiliation{Beijing National Laboratory for Condensed Matter Physics and Institute of Physics, Chinese Academy of Sciences, Beijing 100190, China}
\affiliation{School of Physical Sciences, University of Chinese Academy of Sciences, Beijing 100049, China}
\affiliation{Beijing Academy of Quantum Information Sciences, Beijing, China}

\author{Nandini Trivedi}
\thanks{To whom correspondence should be addressed: txiang@iphy.ac.cn, trivedi.15@osu.edu}
\affiliation{Department of Physics, The Ohio State University, Columbus, Ohio 43210, USA}

%%%%%%%%%%%%

%%%%%%%%%%%%
\begin{abstract}
\begin{center}
    {\textbf{Abstract}}
\end{center}

We propose a mechanism to explain the emergence of an intermediate gapless spin liquid phase in the antiferromagnetic Kitaev model in an externally applied magnetic field, sandwiched between the well-known gapped chiral spin liquid and the gapped partially polarized phase.
We propose that, in moderate fields, $\pi$-fluxes nucleate in the ground state and trap Majorana zero modes. As these fluxes 
proliferate with increasing field, the Majorana zero modes overlap creating an emergent $\mathbb{Z}_2$ quantum Majorana metallic state with a `Fermi surface' at zero energy. We further show that the Majorana spectral function captures the dynamical spin and dimer correlations obtained by the infinite Projected Entangled Pair States method, thereby validating our variational approach. 
%{\color{red}The emergence of the IGP as a Majorana metal at zero temperature indicates a new class of gapless QSLs alongside the commonly recognized Dirac spin liquids and $U(1)$ spinon Fermi surfaces in prevailing theories, bringing new insights into the nature of various candidate QSL phases of matter stabilized by moderate magnetic fields. }
% We discuss the implications of our results for candidate Kitaev materials. 

\end{abstract}
%%%%%%%%%%%%

%%%%%%%%%%%%
\maketitle

\let\oldaddcontentsline\addcontentsline
\renewcommand{\addcontentsline}[3]{}
%%%%%%%%%%%%
\noindent\textbf{Introduction}

Quantum spin liquids (QSLs) are exotic topological quantum matter that transcend the traditional framework of Landau's symmetry-breaking theory. Beyond the absence of zero-temperature order, QSLs are positively characterized by fractionalized degrees of freedom and associated emergent gauge fields that globally constrain the dynamics of these fractionalized particles~\cite{Wen1990,wen2002quantum,kitaev2006anyons,Wen2010,balents2010spin,savary2016quantum,KnolleARCMP2019,ZhouRMP,KHATUA20231}.
With the discovery of the exactly solvable QSLs in Kitaev honeycomb model \cite{kitaev2006anyons,Tao07}, 
recent years have seen significant efforts towards theoretically understanding and experimentally searching for novel QSL phases
in candidate Kitaev materials due to spin-orbital coupling~\cite{Jackeli2009}. Experimental focus has primarily been on the iridate magnetic insulators
A$_2$IrO$_3$ (A = Na, Li and Cu)~\cite{Jackeli2010,2010Na2IrO3,2011Na2IrO3,2015LiIrO3,2017CuIrO3} and $\alpha$-RuCl$_3$  \cite{Knolle2018,Trebst2022,banerjee2017neutron,ywq2017}. 
Recently, Na$_3$Ni$_2$BiO$_6$~\cite{Shangguan2023}, Na$_2$Co$_2$TeO$_6$~\cite{Lin2021,Yao2022,Pilch23,Gaoting24,chen2024planar} and YbOCl~\cite{YbOCl2022,YbOCl2024} have emerged as promising candidates hosting antiferromagnetic Kitaev interactions, thereby broadening the scope of research in the quest for QSLs.

\begin{figure}[t]
\centering
\includegraphics[width=0.95\columnwidth]{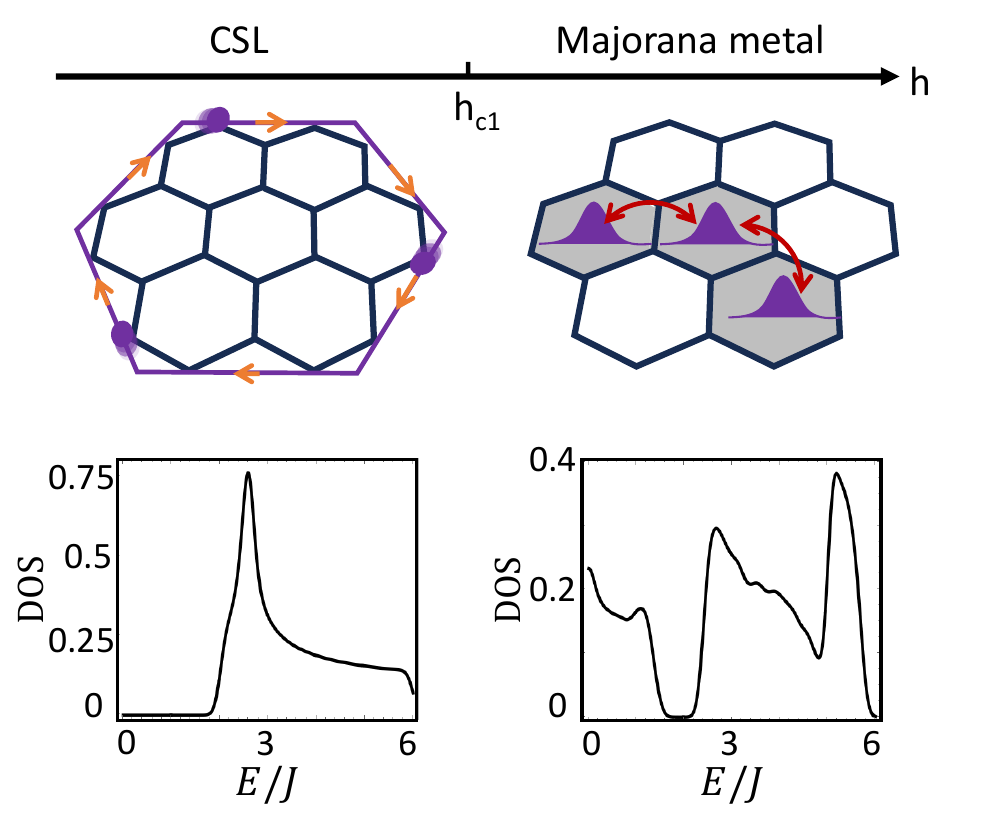}
\caption{Emergence of the Majorana metal from the chiral spin liquid (CSL) under moderate magnetic fields at a critical field $h_{c1}=0.45J$. The first row depicts a schematic of the CSL and Majorana metal phases, where the dots in the left plot represent boundary chiral Majorana modes while the Gaussian-like wavepackets in the right plot indicate Majorana zero modes trapped at $\pi$-fluxes. The second row shows our results for their respective bulk density of states (DOS) under periodic boundary conditions. The two DOS are calculated from the Majorana-hopping models, one without sign disorder ($\overline{W}_{p}=1$, left) and the other with ($\overline{W}_{p}=0.05$, right).  In both cases we used $\lambda=0.25$ for next neighbor hopping in the Majorana hopping model. 
}
\label{fig:TItometal}
\end{figure}

The integrable Kitaev honeycomb model is known to harbor a Dirac QSL in the fractionalized quantum sector of Majorana fermions, which becomes a chiral spin liquid (CSL) under time-reversal-breaking perturbation. Remarkably, beyond the Dirac and CSL phases, recent numerical
studies have unveiled a novel gapless phase emerging from antiferromagnetic Kitaev honeycomb model under moderate magnetic field  \cite{David2019,Patel12199,Hu2024}, which is experimentally relevant as QSL phases in candidate materials like Na$_3$Ni$_2$BiO$_6$, Na$_2$Co$_2$TeO$_6$ and YbOCl are often stabilized under moderate magnetic fields. Despite lots of numerical efforts to understand this intermediate gapless phase (IGP) \cite{Patel12199,hickey2019emergence,Jiang_arXiv_2018,Zhu_PRB_2018,Gohlke_PRB_2018,gohlke2018dynamical,jhk21,Baskaran2023,Lihan24,wang2024},
the mechanism underlying its emergence, as well as the nature of its fractionalization and emergent gauge structure, remain elusive and subject to ongoing debate. Inspired by recent numerical studies revealing the importance of fluctuations of $\mathbb{Z}_2$ fluxes in the formation of the IGP \cite{Baskaran2023,wang2024}, we develop a mean-field ansatz for the emergence of a Majorana metallic phase from the gapped CSL phase due to the field-induced proliferation of $\mathbb{Z}_2$ fluxes. In contrast to the thermal Majorana metal induced by thermal fluctuations as discussed in previous literature~\cite{Knolle2019}, %demonstrated that thermal fluctuations can drive the CSL phase to become a Majorana metal. 
in our model it is the quantum fluctuations stemming from the hybridization of fluxes and Majoranas that leads to the quantum phase transition from a CSL to a quantum Majorana metal.
Specifically, we analyze the entanglement between itinerant Majorana fermions and
localized $\mathbb{Z}_2$ fluxes in the IGP, and show the following:
(1) In the background of fluctuating $\mathbb{Z}_2$ fluxes, the massive Majorana fermions of the gapped CSL become metallic in the IGP [Fig.~\ref{fig:TItometal}], and the low-energy fractionalized Majorana fermions couple to $\mathbb{Z}_2$ gauge fields instead of complex fermions with U(1) gauge fields. 
(2) The emergent Majorana metal has a finite `Fermi surface' (FS) at zero energy which evolves as a function of the magnetic field in the IGP.
(3) The dynamical spectral functions of two- and four-spin correlators decomposed in terms of multi-Majorana correlators in the background of fluctuating fluxes is found to agree well with the results by iPEPS obtained with fine energy-momentum resolution. Although our argument is based on the specific model, it demonstrates a novel and general mechanism for the formation of a $\mathbb{Z}_2$ neutral FS in chiral QSLs. The existence and the nature of the emergent IGP as a $\mathbb{Z}_2$ Majorana metal at zero temperature establish a new class of gapless QSLs alongside those commonly recognized, such as U(1) Dirac QSLs and U(1) spinon Fermi surfaces in prevailing theories. It is hence significant not only theoretically but also in relation to experimental observations of gapless quantum spin liquids.

\bigskip
\noindent\textbf{Results}

\noindent \magenta{\it Field-induced Majorana Metal.--} 
We begin with the isotropic Kitaev model on a honeycomb lattice,  with a magnetic field applied normal to the honeycomb plane:
\begin{align}
    H = \sum_{\langle ij\rangle_\alpha} J{\sigma^\alpha_i \sigma^\alpha_j} - h\sum_{i,\alpha}\sigma_i^\alpha ,~~\alpha\in \{x,y,z\}
    \label{eq_ham}
\end{align}
where $\expval{ij}_\alpha$ denotes  nearest-neighbor sites on an $\alpha$-type bond. We show that under a moderate magnetic field normal to the honeycomb plane, the gapped chiral spin liquid (CSL) phase of the Kitaev honeycomb model transitions to the IGP, described as a \emph{neutral bulk superconducting Majorana metal with a finite FS at zero energy}.

%%%%%%%%%%%%%%%%%%%%%%%%%%%
%%%%%%%%%%%%%%%%%%%%%%%%%%%
Equation~\eqref{eq_ham} with $h=0$ can be exactly solved by fractionalizing the spin into two distinct quantum sectors: the gapless Dirac Majorana fermions and the gapped $\mathbb{Z}_2$ fluxes  \cite{kitaev2006anyons}. The corresponding ground state has been shown to have no flux and accommodates only the Dirac Majorana fermions ~\cite{Lieb1994}.
In the regime of weak magnetic fields, small perturbations do not close the gap ($\sim 0.26 J$) of the $\mathbb{Z}_2$ flux excitation, and thus the $\mathbb{Z}_2$ flux sector is still dominated by the vacuum state $\ket{\mathcal{F}}_{0}$ with no flux excitation. A third-order perturbation provides the Hamiltonian $H_{\mathcal{M}}=\bra{\mathcal{F}_{0}}H\ket{\mathcal{F}_{0}}$ in the Majorana sector:
\begin{equation}
\label{eq:Majoranahopping}
H_{\mathcal{M}}=\sum_{j,k} i t_{jk}c_{j}c_{k} + {\rm H.c.}, 
\end{equation}
where $t_{jk}=J u_{jk}$ for nearest-neighbor hoppings; and $t_{jk}=\lambda u_{jl}u_{lk}$ for anti-clockwise next-nearest-neighbor (NNN) hoppings inside hexagons, with $\lambda \propto h^3$ being the leading-order perturbation coefficient. $u_{jk}\equiv \bra{\mathcal{F}_{0}}\hat{u}_{ij}\ket{\mathcal{F}_{0}}$ in Eq.~\eqref{eq:Majoranahopping} is the expectation value of the $Z_{2}$ vector gauge potential $\hat{u}_{jk}$ defined on the bond connecting sites $j$ and $k$. The gauge invariant flux operator corresponding to the flux excitation is the product of six link operators $\hat{u}_{jk}$ that belong to a hexagon, i.e., $\hat{W}_{p}=\prod_{\varhexagon}\hat{u}_{ij}$. The flux-free vacuum state has $\bra{\mathcal{F}_{0}}\hat{W}_{p}\ket{\mathcal{F}_{0}}=+1$ for every hexagon. 
Since the ground state of the model is flux-free \cite{Lieb1994}, and $\hat{u}_{jk}$ is a good quantum number in Eq.~\eqref{eq:Majoranahopping}, we can choose the gauge to be $u_{jk} = 1$ for every bond for the ground state.
This Majorana-hopping model captures the gapped CSL phase characterized by a nonzero Chern number in Majorana bands. As the field strength increases to a moderate level, there is a phase transition to an IGP based on our previous simulations \cite{David2019,Patel12199}, where the simple free Majorana model no longer applies.

We discuss below the mechanism by which the IGP emerges under moderate field, largely from the interplay between flux fluctuations and the Majorana Chern band. We start by establishing a suitable mean field ansatz that captures the essence of the many-body tensor network representation \cite{wang2024} within a quasi-particle picture. Since the Majoranas and $Z_{2}$ fluxes form a complete basis for the Hilbert space, we can write an ansatz for IGP:
\begin{equation} \label{eq:mft}
    \ket{\Psi_{\text{IGP}
    }} = \sum_{\mathcal{F}} \psi_\mathcal{F} \ket{\mathcal{F}} \otimes \ket{\mathcal{M}_\mathcal{F}},
\end{equation}
where $\ket{\mathcal{F}}$ denotes a state that corresponds to a disordered flux configuration on the honeycomb lattice; $\ket{\mathcal{M}_\mathcal{F}}$ denotes the Majorana state conditioned on the flux configuration; and $\psi_\mathcal{F}$ a complex scalar conditioned on $\mathcal{F}$. Once we average over all the flux configurations
$\ket{\mathcal{F}}$ we recover a translationally invariant $\ket{\Psi_{\rm IGP}}$. This is because all flux patterns obtained by translating  $\ket{\mathcal{F}}$ appear in the linear combinations with the same coefficient. 
We propose that $\ket{\mathcal{M}_{\mathcal{F}}}$ is the ground state of $H_{\mathcal{M}}= \bra{\mathcal{F}}H\ket{\mathcal{F}}$, and $H_{\mathcal{M}}$ is given by Eq.~\eqref{eq:Majoranahopping} with sign-disorder in $u_{ij}$. 
We note that the ansatz in Eq.~\eqref{eq:mft} is distinct from existing microscopic MFTs that attempt to explain the origin of the IGP in moderate fields by solving self-consistent equations of quadratic partons~\cite{Jiang2020,ZhangNatComm2022}, which fall short in representing the entanglement between the two fractionalized quantum sectors, and in representing a flux as a physical degree of freedom with many-body entanglement among the six $\hat{u}_{ij}$'s of a hexagon. More justifications for our ansatz can be found in Supplementary Information (SI). 

To understand the origin of the sign disorder in $u_{ij}$, we trace out the flux sector and get the density matrix of Majorana:
\begin{equation}
    \rho_\mathcal{M} = \Tr_{\mathcal{F}} \ket{\Psi_{\text{IGP}}}\bra{\Psi_{\text{IGP}}} = \sum_{\{\mathcal{F}\}}\abs{\psi_\mathcal{F}}^2 \ket{\mathcal{M}_\mathcal{F}}\bra{\mathcal{M}_\mathcal{F}}.
\end{equation} 
Unlike the CSL phase, the flux fluctuation is significant in IGP, and $|\psi_{\mathcal{F}}|^2$ is typically nonzero because of the presence of a large number of plaquette fluxes. 
We assume the flux sector has slow dynamics in the IGP regime according to recent numerical results \cite{Baskaran2023}, so that for a given flux configuration $\ket{\mathcal{F}}$, $H_{\mathcal{M}}$ is determined by the Eq.~\eqref{eq:Majoranahopping} with $\{u_{ij}=\bra{\mathcal{F}}\hat{u}_{ij}\ket{\mathcal{F}}\}$. $H_{\mathcal{M}}$'s conditioned on different flux configurations form an ensemble. 
We emphasize that configurations $\{u_{ij}\}$ related by gauge transformations are physically equivalent, as they produce identical flux patterns, ensuring that all physical quantities remain unchanged.
By sampling the gauge fields $u_{ij}$, we include these slow gauge fluctuations as ``flux disorder" seen by the Majorana sector; after averaging over all random flux patterns in the ensemble, translation symmetry is recovered, and we then obtain the momentum-resolved spectral function for the Majorana sector. 

The strength of the external field enters $H_{\mathcal{M}}$ by determining the strength of NNN hopping $\lambda$ and affecting the density of the $\pi$-fluxes given by $\{u_{ij}\}$. We define the ensemble average of the flux to be $\overline{W}_{p}\equiv \sum_{\mathcal{F}}|\psi_{\mathcal{F}}|^2 W_{p}^{\mathcal{F}}$ with $W_{p}^{\mathcal{F}}=N_{\varhexagon}^{-1}\sum_{\varhexagon}\bra{\mathcal{F}}\hat{W}_{p}\ket{\mathcal{F}}$. $N_{\varhexagon}$ is the number of hexagons in the system. Since a stronger magnetic field can induce a higher density of $\pi$-fluxes, $\overline{W}_{p}$ decreases as the external magnetic field increases. Therefore, for stronger magnetic field, we have an ensemble of $H_{\mathcal{M}}$ with smaller $\overline{W}_{p}$ within the IGP phase. Given the translation symmetry and the three-fold rotation symmetry of our ansatz $\ket{\Psi_{\rm IGP}}$, the expectation of $u_{ij}$ and $W_{p}$ should be identical on each bond and within each hexagon respectively, i.e., each $u_{ij}$ and $W_{p}$ have the same probability to flip. For this reason,
we implement the sign disorder in the gauge fields by randomly flipping each $u_{jk}$ independently with a given probability. Note that similar implementation has been used to study the thermal Majorana metal phase at high temperature in Kitaev's honeycomb model, yielding results that align perfectly with quantum Monte Carlo simulations~\cite{yoshitake2017}. In our implementation of sign-disorders, $W_{p}^{\mathcal{F}}$ of the ensemble $\{\ket{\mathcal{F}}\}$ follows a Gaussian-like distribution with the mean $\overline{W}_{p}$ due to the central limit theorem; for 
details see SI. Our analysis of the Majorana spectrum from Eq.~\eqref{eq:Majoranahopping} upon ensemble averaging reveals an entrance into a gapless Majorana metal from a gapped CSL, as illustrated in Fig.~\ref{fig:TItometal}. While we have utilized a specific scheme to generate the emergent flux disorder, we emphasize that the emergence of the gapless Majorana metal phase does not rely on the specific distribution. Another way to generate flux disorders is discussed in SI to support this point. The underlying mechanism is more general, as we discuss below.

\begin{figure}[t]
\centering
\includegraphics[width=1\columnwidth]{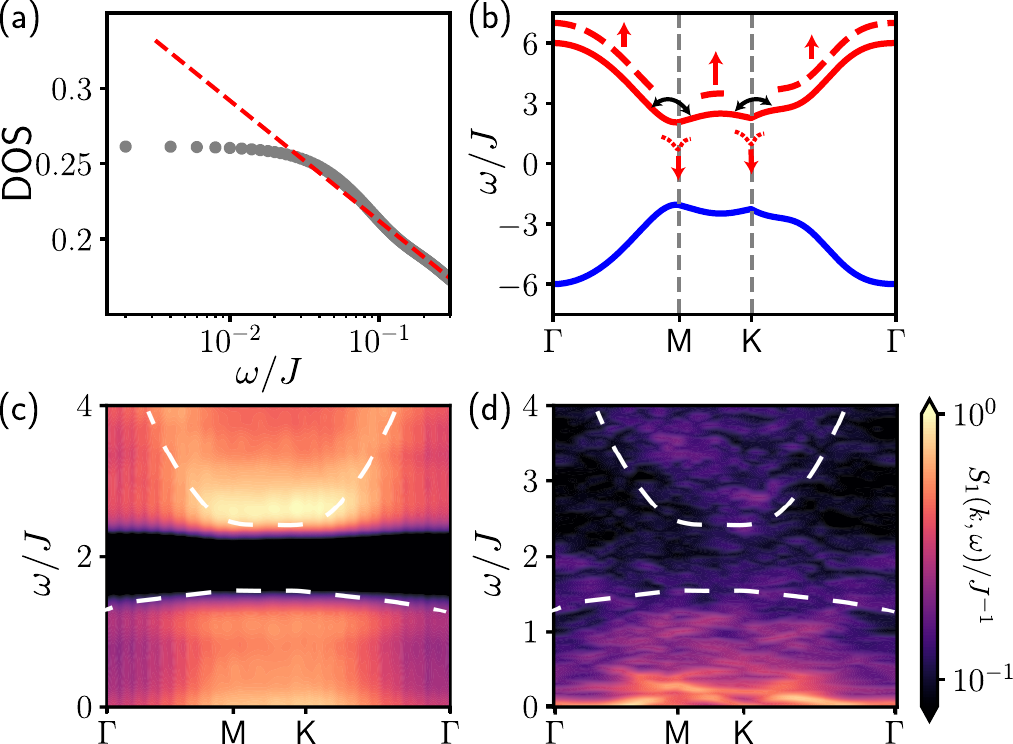}
\caption{Nature of intermediate gapless phase (IGP): (a) The DOS in the Majorana metal shows a logarithmic scaling behavior explaining the origin of states at zero energy. The data are obtained for a system with $\lambda=0.25$,  $80\times 80$ unit cells, averaging over 20 samples with $\overline{W}_{p}=0.05$. (b) Schematic illustration of a minigap opening
due to scattering between Majorana Bloch states induced by fluctuating flux disorder. (c) The spectral function (averaged over 50 samples) of itinerant Majoranas is shown along a high symmetry cut for the IGP, which matches well with iPEPS results for spin-spin correlations, $S_{1}(\mathbf{k},\omega)$, displayed in (d). The white dashed lines in (c) and (d) are eye-guiding lines that emphasize the shape of the spectra and the presence of the gap.
\label{fig:Majoranametal} }
\end{figure}

To understand the transition in the MFT, we first remind that Eq.~\eqref{eq:Majoranahopping} with $u_{ij}=1$ for all bonds is equivalent to a $p+ip$ topological superconductor, as detailed in SI, in which a $\pi$-flux traps a Majorana zero mode (MZM). When the flux density is large enough such that the average separation between two fluxes is comparable to the localization length of the MZMs, then the MZMs can tunnel from one trapped location to the next and form a band around zero energy. Besides the obvious nonzero density of states (DOS) at zero energy shown in Fig.~\ref{fig:TItometal}, we also obtain the DOS as a function of low energies shown in Fig.~\ref{fig:Majoranametal}(a), exhibiting a $\operatorname{ln}|E|$ scaling as expected for a Majorana metal~\cite{Senthil2000,Huse2012,Zhu21,wang2024}. Since a larger gap in the CSL phase results in a smaller localization length for MZMs, we expect that a larger field will be required to enter the IGP, see SI. 
 
While one can understand the gapless nature of IGP from the flux-induced proliferation of MZM and the $\ln |E|$ scaling of DOS near zero energy, to establish that the IGP is indeed metallic with delocalized states, we analyze the scaling of the inverse participation ratio (IPR). The IPR analysis further reveals the multifractal nature of the zero energy states, and from the behavior at large system size we establish the extended nature of these states, in spite of the presence of flux disorder; see SI.

Our results are consistent with the fact that the random Majorana-hopping models of class D are known to exhibit three distinct phases: a topological insulator, a trivial insulator, and a gapless metal phase dubbed as Majorana metal. The topological insulator phase corresponds to the CSL phase of the Kitaev honeycomb model, where the band of Majorana fermions possesses a nonzero Chern number. In this work we identify the IGP in the Kitaev honeycomb model under a moderate magnetic field as a quantum Majorana metal phase at zero temperature. The entrance into the Majorana metal phase from the CSL phase can be understood from another useful perspective: The states in the Majorana sector of the CSL phase can be scattered by the fluctuations of the flux configurations between states with different momenta. This results in a mini-gap, as schematically depicted in Fig.~\ref{fig:Majoranametal}(b) around $\omega \sim 2J$, pushing down the states at lower energy and lifting up the states at higher energy, as seen in MFT and also validated by iPEPS data shown in Fig.~\ref{fig:Majoranametal}(c,d). 
In the absence of flux disorder, the band bottom is pinned at $\omega=2J$ at the $M$ point for large fields $h$; see SI. This is because increasing $h$ enhances the second-nearest-neighbor hopping $\lambda\propto h^3$ which increases the gap at the $K$ point, leaving the energy unchanged at the $M$ point. Consequently, under a sufficiently strong magnetic field, the band bottom remains at $M$ at $\omega = 2J$, where a minigap forms in the IGP phase.
This picture helps us understand the momentum-resolved spectral function and dynamical structures at low energy for the IGP as discussed below.

\bigskip

\noindent \magenta{\it Spin dynamics.--} 
We start by writing the time-dependent spin-spin correlations in terms of the Majorana and flux states, 
\begin{equation}
\begin{aligned}
\label{eq:spincorr}
    \expval*{\boldsymbol{\sigma}_i(t)\cdot \boldsymbol{\sigma}_j(0)}
    \sim \langle\bra{\mathcal{M}_\mathcal{F}} c_i(t) c_j\ket{\mathcal{M}_\mathcal{F}}\rangle_{\mathcal{F}},
\end{aligned}
\end{equation}
 of which the Fourier transformation is denoted as $S_{1}(\mathbf{k},\omega)$. On the left-hand side $\expval{\bullet} \equiv \bra{\Psi_{\text{IGP}}} \bullet \ket{\Psi_{\text{IGP}}}$, on the right side the outer bracket denotes disorder average over flux configurations. 
We apply the ansatz in Eq.~\eqref{eq:mft} assuming the flux dynamics is much slower than the Majorana dynamics~\cite{Baskaran2023}. Hence the dynamical spin spectrum is determined primarily by Majorana fermions in the diagonal sector (see SI).
For $i,j$ belonging to the same sublattice of the honeycomb lattice, the Fourier transform of $\langle \bra{\mathcal{M}_\mathcal{F}} c_i(t) c_j\ket{\mathcal{M}_\mathcal{F}}\rangle_{\mathcal{F}}$ gives the average of single-particle Majorana spectrum over all flux configurations (see SI), providing distinct features of the emergent Majorana FS.

%%%%%%%%%%%%%%%
\begin{figure}[t]
    \centering
    \includegraphics[width=0.99\linewidth]{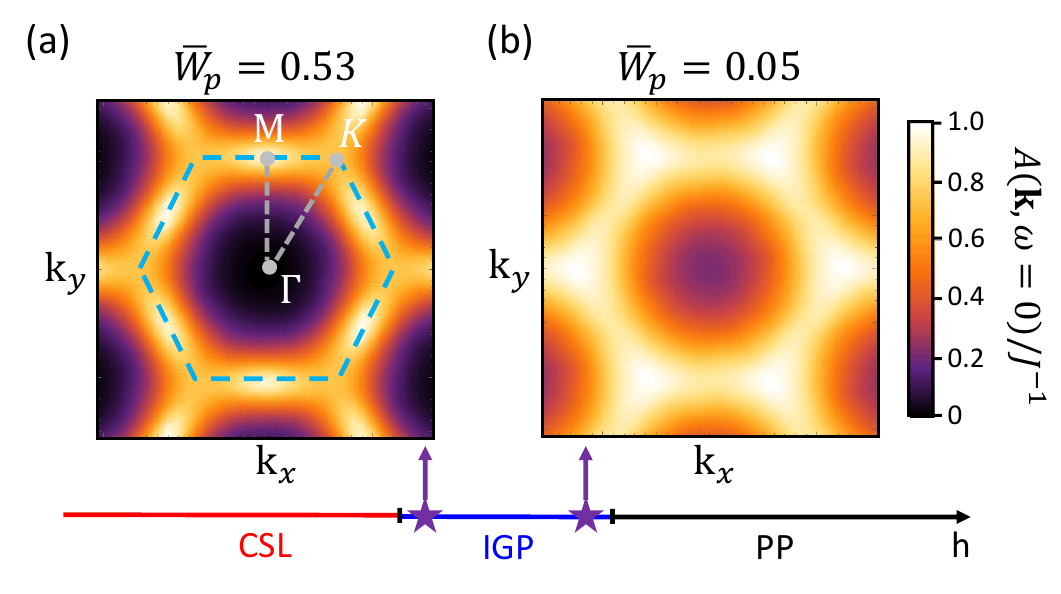}
    \caption{
    Majorana spectral function at zero energy on the FS, $A(\mathbf{k},\omega=0)$, in the intermediate gapless phase (IGP) sandwiched between the chiral spin liquid (CSL) phase and the polarized phase (PP), for (a) $\overline{W}_{p}=0.53$ and (b) $\overline{W}_{p}=0.05$, where $\overline{W}_{p}$ is the ensemble average of the $\pi$-flux density. $h$ represents the strength of the Zeeman field [c.f. Eq.~\eqref{eq_ham}]. The blue dashed contour in (a) marks the boundary of the first BZ and the gray dashed lines represent the cut used in Fig.~\ref{fig:Majoranametal}. $\lambda=0.25$ is used for the calculations.  }
    \label{fig:singlespin}
\end{figure}
%%%%%%%%%%%%%%%

We now test Eq.~\eqref{eq:spincorr} by comparing the single-particle Majorana spectrum to the spin-flip dynamics obtained by iPEPS. 
The Majorana spectral function is computed by:
\begin{equation}
\label{eq:specfunc}
    \begin{split}
        A(\mathbf{r}_1,\mathbf{r}_2,\omega)
        = \sum_n \operatorname{Tr}_{\text{cell}}\phi_n(\mathbf{r}_1) \phi_n^*(\mathbf{r}_2) \delta(\omega - E_n)
    \end{split}
\end{equation}
where $\phi_n(r) = \braket{r}{n}$, with $\ket{n}$ being an eigenstate of $H_{\mathcal{M}}$. Note that $\operatorname{Tr}_{\text{cell}}$ is the trace of intracell degrees of freedom, which corresponds to the sum of intra-sublattice correlations whose momentum-space representation is periodic within the first Brillouin zone (BZ). 
Since the ground state of IGP is translation invariant, all flux configurations in the ansatz should form a representation of the translation symmetry. Therefore, there is a well-defined momentum resolved spectral function $A(\mathbf{k},\omega)$ that is the center-of-mass average of $A(\mathbf{r}_{1},\mathbf{r}_{2},\omega)$. This center-of-mass average is equivalent to an average over flux configurations connected by translations, within the average over all flux configurations.  As all the Bloch eigenstates in the clean limit form an orthonormal basis, we expand eigenstates of a disordered system into  $\ket{n}=\sum_{\alpha\mathbf{k}}c^{n}_{\alpha\mathbf{k}}\ket{\alpha\mathbf{k}}$, where $\alpha$ is the band index and $\mathbf{k}$ the momentum of the Bloch states. Substituting this into Eq.~\eqref{eq:specfunc} and averaging over center-of-mass $\mathbf{R}=(\mathbf{r}_1+\mathbf{r}_2)/2$, we have
\begin{equation}
\label{eq:specfuncave}
 \begin{split}
       \langle A(\mathbf{r},\mathbf{R},\omega)\rangle_{\mathbf{R}} &=\frac{1}{N}\sum_{n\alpha\mathbf{k}}e^{i\mathbf{k}\cdot\mathbf{r}}|c^{n}_{\alpha\mathbf{k}}|^2\delta(\omega - E_n),
    \end{split}
\end{equation}
where $\mathbf{r}=\mathbf{r}_{1}-\mathbf{r}_2$. After a Fourier transformation, one can get the spectral function in momentum space:
\begin{equation}
\label{eq:specfunck}
 \begin{split}
A(\mathbf{k},\omega)&=\sum_{n\alpha}|c^{n}_{\alpha\mathbf{k}}|^2\delta(\omega - E_n).
    \end{split}
\end{equation}
Next, we average Eq.~\eqref{eq:specfunck} over random flux configurations that are not connected by translations.
We show our results of averaged $A(\mathbf{k},\omega)$ along high-symmetry lines in Fig.~\ref{fig:Majoranametal}(c), which captures the gapless feature of the spin-spin correlations around $\rm M$ and $\rm K$. 
For comparison we also show numerical evidence from iPEPS for intra-sublattice spin-spin correlations in Fig.~\ref{fig:Majoranametal}(d); details can be found in SI.  Remarkably, our mean field analysis when compared with the unbiased iPEPS results, reproduces the two most salient features in the IGP: (i) The presence of spectral weight at low energies immediately above $\omega = 0$; and (ii) the opening of a gap centered around $\omega \simeq 2 J$, which we now understand as a flux-induced gap separating the upper and the lower branches of the Majorana bands.

To gain more information about the gapless modes in the Majorana metal, 
we further calculate the averaged $A(\mathbf{k},\omega=0)$ over the whole BZ (i.e., the FS) for systems with different $\overline{W}_{p}$. 
The results in Fig.~\ref{fig:singlespin} explicitly show the presence of gapless states around $\rm M$ and $\rm K$ in the Majorana sector, which suggests the presence of zero-energy modes with definite momentum or a Majorana FS.
It is straightforward to see that when $\overline{W}_{p}$ is large (sparse fluxes) for weak magnetic fields, zero energy states are mainly found around the $\rm M$ point. However, when $\overline{W}_{p}$ becomes smaller (denser fluxes) for stronger magnetic fields, zero energy states populate mainly around $\rm K$ points.  This matches well with our previous iPEPS calculations~\cite{wang2024}. The reason zero energy states first appear near the $\rm M$ point before appearing near the $\rm K$ point is because in the clean limit the band edge is at the $\rm M$ point for large enough NNN hopping $\lambda$, as depicted in Fig.~\ref{fig:Majoranametal}(b), and is thus more susceptible to be rendered gapless by fluxes. More information about the band structures for different $\lambda$'s can be found in SI. 
It is important to distinguish the fluctuations of flux configurations that has averaged translation symmetry considered here from quenched bond- or site-vacancy disorder discussed in Ref.~\cite{Knolle19,Kao21}. While both types of disorder can lead to gapless states, quenched disorder lacks well-defined momenta for zero-energy modes due to the absence of translation symmetry. 

We briefly comment on the implications of the Majorana FS on quantum oscillations. There have been previous attempts to attribute the quantum oscillations to a gapless U(1) spin liquid phase characterized by a Fermi surface of complex fermions \cite{Jiang_arXiv_2018,hickey2019emergence,Patel12199,czajka2021oscillations}. However, as discussed in Ref.~\cite{baskaran2015majorana}, even if interactions (such as pairing between U(1) fermions) convert the complex fermions into real (Majorana) fermions, the de Haas van Alphen quantum oscillations can still be observed as long as some fraction of Majorana fermions remaining gapless. Similarly, Friedel-type oscillations can also occur due to the presence of a FS of U(1) complex fermions and can persist under $\mathbb{Z}_2$ gauge with fermion pairing~\cite{Kohsaka24,zhang2023machine,fetter1965spherical}.
 
\bigskip

\noindent \magenta{\it Dimer dynamics.--}
Similar to the spin dynamics, the dimer correlations reflect the dynamics of four-Majorana correlations that can be observed by Resonant Inelastic X-ray Scattering (RIXS) experiments~\cite{Kumar2018,Schlappa2018}. We denote the dimer by $\mathcal{D}_j^\alpha = \sigma_{j}^\alpha \sigma_{j+z}^\alpha$. In analogy to the two-spin spectrum, the dimer-dimer spectrum $S_2 (\mathbf{k},\omega) = {\rm F.T.}\{\expval*{\boldsymbol{\mathcal{D}}_i(t) \cdot \boldsymbol{\mathcal{D}}_j}\}$ in the IGP can be approximated by:
\begin{equation}
    \expval*{\boldsymbol{\mathcal{D}}_i(t) \cdot \boldsymbol{\mathcal{D}}_j} \sim \langle\bra{\mathcal{M}_\mathcal{F}} c_{i}(t) c_{i+z}(t) c_j c_{j+z}\ket{\mathcal{M}_\mathcal{F}}\rangle_{\mathcal{F}} \label{eq:dimer}
\end{equation}
An exact evalutaion of the R.H.S. of Eq.~\eqref{eq:dimer} using mean-field modes is not entirely justified since the Majorana dispersion is rather broad indicating strong lifetime effects (see Fig.~\ref{fig:Majoranametal}(c)). We therefore attempt to capture the essence using a single-mode approximation (SMA) to describe the excitations of the gapless continuum of the Majorana fermions near low energy. Such a semi-quantitative description of the Majorana FS remarkably shows good agreement with the data obtained by iPEPS.
\begin{figure}[t]
    \centering
    \includegraphics[width=\linewidth]{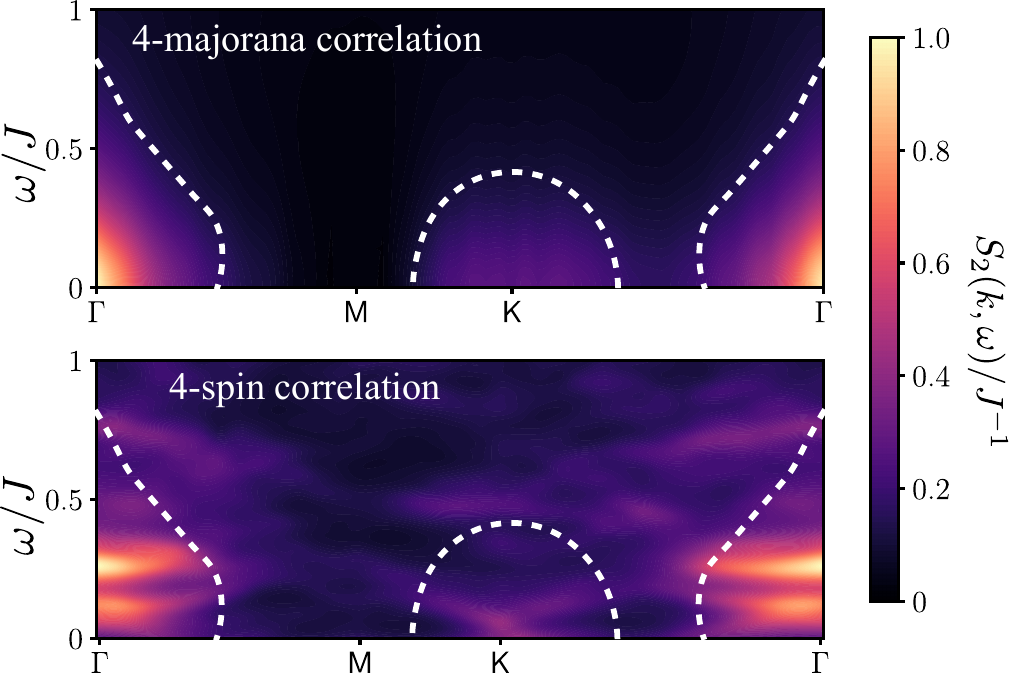}
    \caption{Comparison between the four-Majorana correlation in Eq.~\eqref{eq:s2zz} (top) and the dimer-dimer correlation in the intermediate gapless phase obtained by iPEPS (bottom) along the momentum cut through high-symmetry points $\Gamma\rm M K \Gamma$. The white dashed lines are eye-guiding lines that enclose the energy-momentum region having the strongest intensities, i.e. those near $\rm K(K')$ and $\Gamma$ at the lowest energy; while signals around $\rm M$ point are negligible. }
    \label{fig:dimer}
\end{figure}

Informed by the spectral function at the Fermi energy presented in Fig.~\ref{fig:singlespin}(b), we use a low-energy SMA around the soft fermion modes at and near the $\rm K$ and $\rm K'$ points, depicting the state deep in the IGP where flux density is near half-filling.  It is then straightforward to calculate the dimer-dimer correlation in terms of the effective Majorana band, which in Lehmann spectral representation is:
\begin{equation} \label{eq:s2zz}
\begin{split}
        S_2 (\mathbf{k},\omega) \simeq \int_{\mathbf{q}\in \rm BZ} W(\mathbf{k}, \mathbf{q}) \delta[\omega - (E_{\mathbf{k} - \mathbf{q}} + E_{\mathbf{q}})]
\end{split} 
\end{equation}
up to constants. Here $W(\mathbf{k}, \mathbf{q}) = E_{\mathbf{k} - \mathbf{q}}^2 / (E_{\mathbf{k} - \mathbf{q}}^2 - Q_{\mathbf{k} - \mathbf{q}}^2)$ is the two-fermion (four-Majorana) spectral weight function given by the approximated single-mode Majorana band $E_{\mathbf{k}}$, gapless near $\rm K$ and $\rm K'$ points in keeping with Fig.~\ref{fig:singlespin}(b); and $Q_\mathbf{k} \sim [\sin(\mathbf{k}\cdot \mathbf{n}_2) - \sin(\mathbf{k}\cdot \mathbf{n}_1) - \sin(\mathbf{k}\cdot (\mathbf{n}_2-\mathbf{n}_1))]$, the NNN hopping induced by a time-reversal-breaking perturbation, concomitantly tuned to give the desired soft modes in the aforementioned $E_{\mathbf{k}}$.
The derivation and computational details are relegated to the SI.
Results by SMA and iPEPS in Fig.~\ref{fig:dimer} 
show a strikingly similar spectrum at low energies. At the lowest energies, the strongest intensities for the dimer-dimer spectral functions are observed near the $\Gamma$ point, followed by slightly weaker signals at $\rm K$, and negligible signal near $\rm M$. Minor differences between the Majorana analysis and iPEPS can be attributed to truncation errors of iPEPS and the loss of higher-energy states in the MFT. 

\bigskip
\noindent\textbf{Discussion}
% \noindent \magenta{\it Conclusions and remarks.--}

In this work we shed light on the field-induced IGP in the Kitaev honeycomb model by introducing an effective tight-binding model that describes the interplay between emergent $\mathbb{Z}_2$ flux-disorder and Majorana Chern-bands. Within our theory, the IGP is a zero-temperature quantum Majorana metal phase characterized by persistent fermion pairing and a neutral FS.  
In previous works, the intermediate phase has been either identified as a gapless phase of neutral spinons with a finite Fermi surface and an emergent U(1) gauge structure~\cite{Jiang_arXiv_2018,hickey2019emergence,Patel12199,zhang2023machine} or identified as a gapped topological order with Chern number $\pm 4$ and an emergent $\mathbb{Z}_2$ gauge structure \cite{Jiang2020,ZhangNatComm2022}. All of these proposals fail to capture the simultaneous presence of a gapless phase and an emergent $\mathbb{Z}_2$ gauge structure observed in our recent numerical study~\cite{wang2024}. It is by considering the interplay between matter and gauge fields that our work accommodates both the gapless and the $\mathbb{Z}_2$ nature of the IGP.

Recent exact diagonalization studies~\cite{hickey2019emergence,ZhangNatComm2022} claim that the intermediate phase is gapless from the behavior of spin correlations. It is further asserted that the specific heat calculations indicate the proliferation of $\pi$-flux at low energy which leads to a $\mathbb{Z}_2$ to U(1) transition in the gauge structure, resulting in a transition from a chiral spin liquid (CSL) to a spinon metal with a spinon Fermi surface. Such statements are difficult to validate since the calculations are based on spin operators on small systems that are not privy to the gauge structure. In addition, it is difficult to understand from a theoretical perspective the $\mathbb{Z}_2$ to U(1) transition in the gauge structure.

In our approach, as discussed above, we have proposed that the $\mathbb{Z}_2$ gauge structure persists under the transition from a gapped CSL to a gapless Majorana metal, and we further present a mechanism for the transition that is supported by features observed in our iPEPS calculations. 
%We would like to contrast our work with existing theories regarding the intermediate phase by commenting specifically on the results from Ref.~\cite{hickey2019emergence} and Ref.~\cite{ZhangNatComm2022}. In Ref.~\cite{hickey2019emergence}, spin correlations calculated via exact diagonlization support the gapless nature of the intermediate phase, and the specific heat calculations indicate the proliferation of $\pi$-flux at low energy. However, the proposal in Ref.~\cite{hickey2019emergence}-- that $\pi$-flux proliferation leads to a $\mathbb{Z}_2$ to $U(1)$ transition in the gauge structure, resulting in a transition from a chiral spin liquid (CSL) to a spinon metal with a spinon Fermi surface -- lacked further evidence and explanation. In contrast, we propose that the $\mathbb{Z}_2$ gauge structure should persist and present a clear mechanism for the transition from a gapped CSL to a gapless Majorana metal, supported by features observed in iPEPS calculations. }

Ref.~\cite{ZhangNatComm2022} presents a microscopic parton mean field theory (MFT) that provides an explanation of the divergent susceptibility at the transition between the CSL and intermediate phase, consistent with previous DMRG results~\cite{Patel12199,Jiang_arXiv_2018}. 
However, the MFT also finds that the intermediate phase is gapped with a low-energy ring of gapped excitations around the $\Gamma$ point in momentum space. This result is however not consistent with our unbiased iPEPS calculations that show a logarithmically divergent density of states at low energy (Fig.~\ref{fig:Majoranametal}(a) and Ref.~\cite{wang2024})
that strongly support a $\mathbb{Z}_2$ gapless Majorana metal.

Our identification of IGP as a zero-temperature Majorana metal suggests novel thermal transport and spin relaxation within this phase \cite{Zhu21}. The understanding gained from the spin-spin and dimer-dimer correlations in terms of multi-Majorana correlations  in Fig.~\ref{fig:dimer} provides detailed predictions for observation of fractionalization in Raman scattering and momentum-resolved RIXS experiments. 
Given that Majorana metal phases can be induced by thermal fluctuations in chiral spin liquids under zero magnetic field as previously reported~\cite{self2019thermally,Fulga20,eschmann2020partial}, along with our theory on the gapless IGP as a $\mathbb{Z}_2$ Majorana metal at zero temperature induced by quantum fluctuations, these results are suggestive enough to warrant further investigation into the comprehensive phase diagram parametrized by temperature and magnetic field for the new class of the $\mathbb{Z}_2$ gapless QSLs, and the unusual transport properties therein.

\bigskip
\noindent\textbf{Methods}

\noindent \magenta{\it Average of the spectral function.--} The center-of-mass averaged spectral function [c.f. Eq.~\eqref{eq:specfuncave}] can be derived from Eq.~\eqref{eq:specfunc} by decomposing the eigenwavefunction $\phi_{n}(\mathbf{r})$ into a linear combination of Bloch states:
\begin{equation}
\label{eq:decomp}
\phi_{n}(\mathbf{r})=\frac{1}{\sqrt{N}}\sum_{\alpha\mathbf{k}} c^{n}_{\alpha\mathbf{k}}e^{i\mathbf{k}\cdot \mathbf{r}}u_{\alpha\mathbf{k}}.
\end{equation}
where $\alpha$ labels internal degrees of freedom, i.e. the sublattice indices. Substituting Eq.~\eqref{eq:decomp} into Eq.~\eqref{eq:specfunc} and replacing $\mathbf{r}_1$ and $\mathbf{r}_2$ by $\mathbf{r}=\mathbf{r}_1-\mathbf{r}_2$ and $\mathbf{R}=(\mathbf{r}_1+\mathbf{r}_2)/2$, we can derive
\begin{equation}
\label{eq:specfuncderive2}
\begin{split}
\langle &A(\mathbf{r}_1,\mathbf{r}_2,\omega)\rangle_{\mathbf{R}} \equiv \frac{1}{N}\sum_{\mathbf{R}}A(\mathbf{r},\mathbf{R},\omega)\\
&=\frac{1}{N^2}\sum_{\mathbf{R}}\sum_{n}\sum_{\alpha\alpha^{\prime}}\sum_{\mathbf{k}\mathbf{k}^{\prime}}e^{i(\mathbf{k}-\mathbf{k}^{\prime})\cdot\mathbf{R}}e^{i(\mathbf{k}+\mathbf{k}^{\prime})\cdot\mathbf{r}/2}
\\
&\ \  \times c^{n}_{\alpha\mathbf{k}}c^{n\star}_{\alpha^{\prime}\mathbf{k}^{\prime}}\operatorname{Tr}_{\text{cell}}(u_{\alpha\mathbf{k}}u^{\dag}_{\alpha^{\prime}\mathbf{k}^{\prime}})\delta(\omega-E_{n})
\\
&=\frac{1}{N}\sum_{n}\sum_{\alpha\alpha^{\prime}}\sum_{\mathbf{k}\mathbf{k}^{\prime}}\delta_{\mathbf{k}\mathbf{k}^{\prime}}e^{i(\mathbf{k}+\mathbf{k}^{\prime})\cdot\mathbf{r}/2}c^{n}_{\alpha\mathbf{k}}c^{n\star}_{\alpha^{\prime}\mathbf{k}^{\prime}}
\\
&\ \  \times \operatorname{Tr}_{\text{cell}}(u_{\alpha\mathbf{k}}u^{\dag}_{\alpha^{\prime}\mathbf{k}^{\prime}})\delta(\omega-E_{n})
\\
&=\frac{1}{N}\sum_{n}\sum_{\alpha\alpha^{\prime}}\sum_{\mathbf{k}}e^{i\mathbf{k}\cdot\mathbf{r}}c^{n}_{\alpha\mathbf{k}}c^{n\star}_{\alpha^{\prime}\mathbf{k}}\delta_{\alpha\alpha^{\prime}}\delta(\omega-E_{n})
\\
&=\frac{1}{N}\sum_{n\alpha\mathbf{k}}e^{i\mathbf{k}\cdot\mathbf{r}}|c^{n}_{\alpha\mathbf{k}}|^2\delta(\omega-E_{n}).
\end{split}
\end{equation}
After a Fourier transformation, we can eventually derive Eq.~\eqref{eq:specfunck} used in our numerical calculations. Specifically, for each flux pattern generated by randomly flipping $u_{ij}$ on each bond, we diagonalize the disordered Majorana-hopping model and calculate the spectral function using Eq.~\eqref{eq:specfunck}. We then average the results for different random flux patterns which leads to the plot in Figs.~\ref{fig:Majoranametal} and ~\ref{fig:singlespin}. More details about spectral functions of different phases can be found in SI.

\noindent \magenta{\it iPEPS calculation.--}The ground state is described as infinite tensor network with translation invariant local tensor $A$:

\begin{equation} 
    \ket{\psi_0} =\vcenter{\hbox{\includegraphics[scale=0.3]{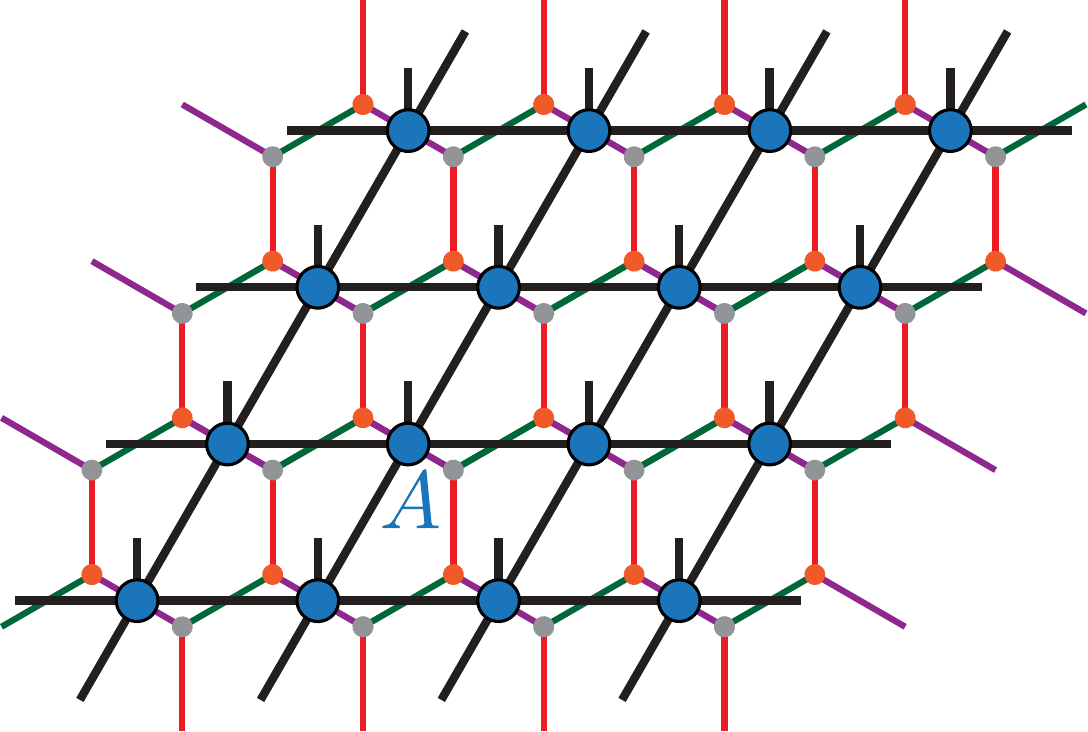}}},
\end{equation}
and tensor $A$ can be derived by minimizing the cost function:
\begin{equation}
    L = {\frac{\bra{\psi_0(A)}H\ket{\psi_0(A)}}{\braket{\psi_0(A)}{\psi_0(A)}}}
\end{equation}
using automatic differentiation \cite{liao2019differentiable}.
The single-mode approximation (SMA), introduced in \cite{feynman1954atomic,haegeman2012variational,vanderstraeten2015excitations,ponsioen2022automatic,xiang2023density}, is employed to characterize excited states. The SMA uses variational tensors for the excited state as shown below:
\begin{equation}
\begin{split}
&\ket{\Psi_\mathbf{k}}=\sum_{\mathbf{r}}e^{-i\mathbf{k}\cdot\mathbf{r}}\ket{\Psi_\mathbf{r}}
\\
&=\sum_{\mathbf{r}}e^{-i\mathbf{k}\cdot\mathbf{r}}\vcenter{\hbox{\includegraphics[scale=0.3]{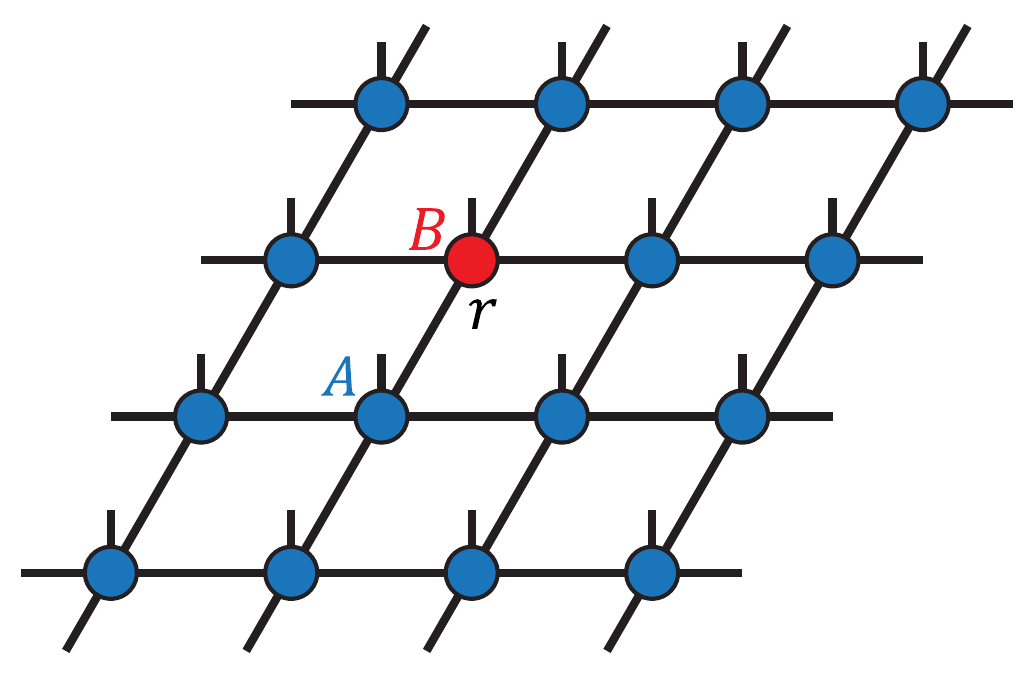}}}
\end{split}
\end{equation}
Here, $\mathbf{k}$ denotes momentum and $\ket{\Psi_\mathbf{r}}$ signifies the state with an excitation at site $\mathbf{r}$.
Under the representation of iPEPS,  it's implemented by replacing the local tensor $A$ of the ground state at site $\mathbf{r}$ with a disturbed local tensor $B$. Note that excited states are required to adhere to the orthogonality constraint in relation to the ground state, depicted as $\braket{\psi_0}{\Psi_\mathbf{k}(B)}=0$. Additionally, they must eliminate gauge redundancy to ensure accuracy in calculations \cite{ponsioen2022automatic}. With these two constraints, we can obtain the $B$ tensor and thus the excited states by minimizing the cost function
\begin{equation}
    L^{\prime} = {\frac{\bra{\Psi_\mathbf{k}(B)}H-E_{\rm gs}\ket{\Psi_\mathbf{k}(B)}}{\braket{\Psi_\mathbf{k}(B)}{\Psi_\mathbf{k}(B)}}}.
\end{equation}

With the ground state and the excited states, we can directly calculate the spin-spin and dimer-dimer correlations as detailed in SI.

%%%%%%%%%%%%%%%%%%%%%%%%%%%%%%%%%%%%%
\section*{Data availability}
%%%%%%%%%%%%%%%%%%%%%%%%%%%%%%%%%%%%%

The data that support the findings of this study are available from the corresponding author upon request.

%%%%%%%%%%%%%%%%%%%%%%%%%%%%%%%%%%%%%
\section*{Code availability}
%%%%%%%%%%%%%%%%%%%%%%%%%%%%%%%%%%%%%

The codes that support the findings of this study are available from the corresponding author upon request.

%%%%%%%%%%%%

%%%%%%%%%%%%Refs
% \bibliographystyle{apsrev}
\bibliography{spectra}
%%%%%%%%%%%%

%%%%%%%%%%%%%%%%%%%%%%%%%%%%%%%%%%%%%
\section*{Acknowledgement}
%%%%%%%%%%%%%%%%%%%%%%%%%%%%%%%%%%%%%
P.Z., S.F. and N.T. are funded by U.S. National Science Foundation's Materials Research Science and Engineering Center under award number DMR-2011876. S.F. also acknowledges support from  Deutsche Forschungsgemeinschaft (DFG, German Research Foundation) under Germany’s Excellence Strategy--EXC--2111--390814868 as well as the Munich Quantum Valley, which is supported by the Bavarian state government with funds from the Hightech Agenda Bayern Plus.
K.W. and T.X. are funded by the National Key Research and Development Project of China (Grants No.~2021ZD0301800) and the National Natural Science Foundation of China (Grants No.~12488201). Authors are grateful to Bruce Normand, Runze Chi, Tong Liu, Adhip Agarwala, Subhro Bhattacharjee, Johannes Knolle, Michael Knap and Chris Laumann for comments and discussions.

%%%%%%%%%%%%%%%%%%%%%%%%%%%%%%%%%%%%%
\section*{Author contributions}
%%%%%%%%%%%%%%%%%%%%%%%%%%%%%%%%%%%%%

P.Z. and S.F. conducted the theoretical study and performed the tight-binding calculations. K.W. performed the iPEPS calculations.  T.X. and N.T. supervised the work. All authors jointly wrote the paper.

%%%%%%%%%%%%%%%%%%%%%%%%%%%%%%%%%%%%%
\section*{Competing interests}
%%%%%%%%%%%%%%%%%%%%%%%%%%%%%%%%%%%%%

The authors declare no competing interests.

\appendix
%--------------------------------------------------------------
%--------------------------------------------------------------
%--------------------------------------------------------------
%--------------------------------------------------------------
%--------------------------------------------------------------
%--------------------------------------------------------------
%%%%%%%%%% Merge with supplemental materials %%%%%%%%%%
\begin{widetext}
\clearpage
\begin{center}
\textbf{\large Supplementary Information for \\ ``Emergent quantum Majorana metal from a chiral spin liquid"}
\end{center}
% \end{widetext}

%%%%%%%%%% Prefix a "S" to all equations, figures, tables and reset the counter %%%%%%%%%%
\setcounter{equation}{0}
\setcounter{figure}{0}
\setcounter{table}{0}
\setcounter{page}{1}
\setcounter{section}{0}
\makeatletter
\renewcommand{\theequation}{S\arabic{equation}}
\renewcommand{\thefigure}{S\arabic{figure}}
\renewcommand{\bibnumfmt}[1]{[S#1]}
%\renewcommand{\citenumfont}[1]{S#1}
%%%%%%%%%% Prefix a "S" to all equations, figures, tables and reset the counter %%%%%%%%%%

\section{Kitaev spin liquid as a p-wave spinon superconductor}
In this section we describe the $p$-wave nature of the Majorana sector of Kitaev honeycomb model. This supports our argument regarding the transition of the Chiral spin liquid from a gapped $p$-wave superconductor to a Majorana metal under flux disorder. It is also a useful picture to describe the dimer-dimer correlations discussed in the main text in terms of amplitude modes of fermions in a spinless superconductor defined on a square lattice.
Based on Kitaev's Majorana decomposition \cite{kitaev2006anyons}, we can construct the complex fermion with canonical fermionic algebra by defining \cite{Baskaran2007}
\begin{equation}
\begin{split}
    c_{i} = f_i^\dagger + f_i,~ c_{i+z} = i(f_i - f_i^\dagger),~
    b_i^\alpha = \chi_{i,\alpha} + \chi_{i,\alpha}^\dagger,~ b^{\alpha}_{i+\alpha} = i(\chi_{i,\alpha} - \chi_{i,\alpha}^\dagger)
\end{split}
\end{equation}
where we assume $i\in A$ sublattice, and $\chi_{i,\alpha}$ being the bond fermion operator on $\alpha$ links.  Under this transformation, the original spin exchanges in the honeycomb model can be expressed as
\begin{equation} \label{eq:trans}
\begin{split}
         \sigma_i^z \sigma_{i+z}^z &= - (2n_i^f - 1)(2n_i^z - 1) \\
        \sigma_i^x \sigma_{i+x}^x &= -(2n_i^x - 1)(f_i f_{i-\delta_1} + f_i^\dagger f_{i-\delta_1} + {\rm H.C.}) \\
        \sigma_i^y \sigma_{i+y}^y &= -(2n_i^y - 1)(f_i f_{i-\delta_2} + f_i^\dagger f_{i-\delta_2} + {\rm H.C.})
\end{split}
\end{equation}
where $n_i^f = f_i^\dagger f_i$, $n_i^\alpha = \chi_{i,\alpha}^\dagger \chi_{i,\alpha}$ for $\alpha=x,y,z$. The exchange in $y$ can be written by the same token as $\sigma_i^x \sigma_{i+x}^x$ under $\delta_1 \leftrightarrow \delta_2,~ x\leftrightarrow y$. In the absence of magnetic field, the bond fermions $\chi$ and complex fermions $f$ live in disjoint Hilbert spaces, hence, in the zero-flux sector, we can fix the gauge by setting $n_i^z = n_i^x = n_i^y = 1$, such that the problem is reduced to a superconducting model defined on a triangular lattice. The generic Hamiltonian, according to Ref.~\onlinecite{feng2023dim}, can be expressed as
\begin{equation}
\begin{split}
    H = -t\sum_{i,j} (f_i^\dagger f_{i+\mathbf{n}_j} + {\rm H.c.} ) 
     + \Delta \sum_{i,j} (f_i^\dagger f_{i+\mathbf{n}_j}^\dagger + {\rm H.c.}) - 2 \mu \sum_i f_i^\dagger f_i
\end{split}
\end{equation}
where $\mathbf{n}_j \in \left\{\mathbf{n}_1 = (-\frac{1}{2},\frac{\sqrt{3}}{2}),\mathbf{n}_2 = \left(\frac{1}{2}, \frac{\sqrt{3}}{2}\right)\right\}$ are two lattice vectors for the triangular lattice, whose reciprocal vectors are $b_1 = (2\pi, 2\pi/\sqrt{3}), ~b_2 = (-2\pi,2\pi/\sqrt{3})$. 
For the isotropic Kitaev model, the coefficients in the flux-free sector are given by
\begin{align}
    t &= J_x (2n_i^x - 1) = J_y(2n_i^y - 1) = J_x\\
    \Delta &= -J_x (2n_i^x - 1) = -J_y (2n_i^y - 1) = -J_x\\
    \mu &= J_z (2n_i^z - 1) = J_z
\end{align}
where we have chosen $n_i^x = n_i^y = n_i^z = 1$ as a translationally invariant gauge for the flux-free condition. In momentum space
\begin{align}
	f_j = \frac{1}{\sqrt{N}} \sum_{k} \exp(ikr_j) f_k,~
	f_j^\dagger = \frac{1}{\sqrt{N}} \sum_{k} \exp(-ikr_j) f_k^\dagger, 
\end{align}
the Hamiltonian then reads
\begin{equation}
	H =  \sum_k (f_k^\dagger,~ f_{-k})
	\begin{pmatrix}
		\xi_k - \mu	 & \Delta_k \\
		 \Delta_k^* & -(\xi_k - \mu) \\
	\end{pmatrix}
	\begin{pmatrix}
		f_k \\ f_{-k}^\dagger
	\end{pmatrix}
\end{equation}
where the dispersion and pairing amplitude are given by
\begin{align}
	\xi_\mathbf{k} = -4t \cos(\frac{k_x}{2}) \cos(\frac{\sqrt{3}k_y}{2}),~
	\Delta_\mathbf{k} = -4 \Delta i \cos(\frac{k_x}{2}) \sin(\frac{\sqrt{3}}{2}k_y).
 \label{eq:pairing}
\end{align}
Supplementary Equation~\eqref{eq:pairing} makes explicit the $p$-wave nature of the low-energy model of $f_{\mathbf{k}}$ ($\Delta_{-\mathbf{k}} = -\Delta_\mathbf{k}$) in the fractionalized quantum sector. 
We next 
% Supplementary Equation~\eqref{eq:Ek} makes explicit the $p$-wave nature of the low-energy model of $f_{\mathbf{k}}$ ($\Delta_{-\mathbf{k}} = -\Delta_\mathbf{k}$) in the fractionalized quantum sector. 
perform a Bogoliubov de Gennes (BdG) transformation parameterized by $u_\mathbf{k}$ and $v_\mathbf{k}$, such that the Hamiltonian becomes diagonal in the new BdG fermionic mode $\gamma_\mathbf{k}$:
\begin{equation}
	\begin{pmatrix}
		\gamma_\mathbf{k} \\ \gamma_{-\mathbf{k}}^\dagger
	\end{pmatrix}
	=
	\begin{pmatrix}
		u_\mathbf{k} & -v_{\mathbf{k}} \\
		-v_\mathbf{-k}^* & u_\mathbf{-k}^*
	\end{pmatrix}
	\begin{pmatrix}
		f_\mathbf{k} \\ f_{-\mathbf{k}}^\dagger
	\end{pmatrix}
\end{equation}
under the unitary constraint $\abs{u_\mathbf{k}}^2 + \abs{v_\mathbf{k}}^2 = 1$. The diagonalization is achieved by 
\begin{equation} \label{eq:uvk}
	\abs{u_\mathbf{k}}^2 = \frac{1}{2} \left( 1 + \frac{\xi_\mathbf{k} - \mu}{E_\mathbf{k}} \right),~~~\abs{v_\mathbf{k}}^2 = \frac{1}{2} \left( 1 - \frac{\xi_\mathbf{k} - \mu}{E_\mathbf{k}} \right)
\end{equation}
with the energy spectrum given by
\begin{align}
    H = \sum_k E_\mathbf{k} \gamma_\mathbf{k}^\dagger \gamma_\mathbf{k} + {\rm const.} ,~
	E_\mathbf{k} = \pm 2\sqrt{(\xi_\mathbf{k} - \mu)^2 + \abs{\Delta_\mathbf{k}}^2} \label{eq:Ek}
\end{align}
The band structure $E_\mathbf{k}$ is presented in Supplementary Figure~\ref{fig:bands}(a). After gapping out the Dirac modes at $\pm \rm K$ by adding time-reversal-(TR)-breaking perturbation $\lambda$, e.g. in Supplementary Figure~\ref{fig:bands}(b), the Hamiltonian can be expanded near $\pm \rm K$ points to reveal the effective $p_x\pm ip_y$ topological superconductor, and the presence of Majorana zero modes trapped in the fluxes, see Ref.~\onlinecite{Read2000}.  
%%%%%%%%%%%%%%%%%%%%%%%%%%%%%%%%%%%%%%%%%%%%%
\begin{figure}[t]
    \centering
    \includegraphics[width=0.9\linewidth]{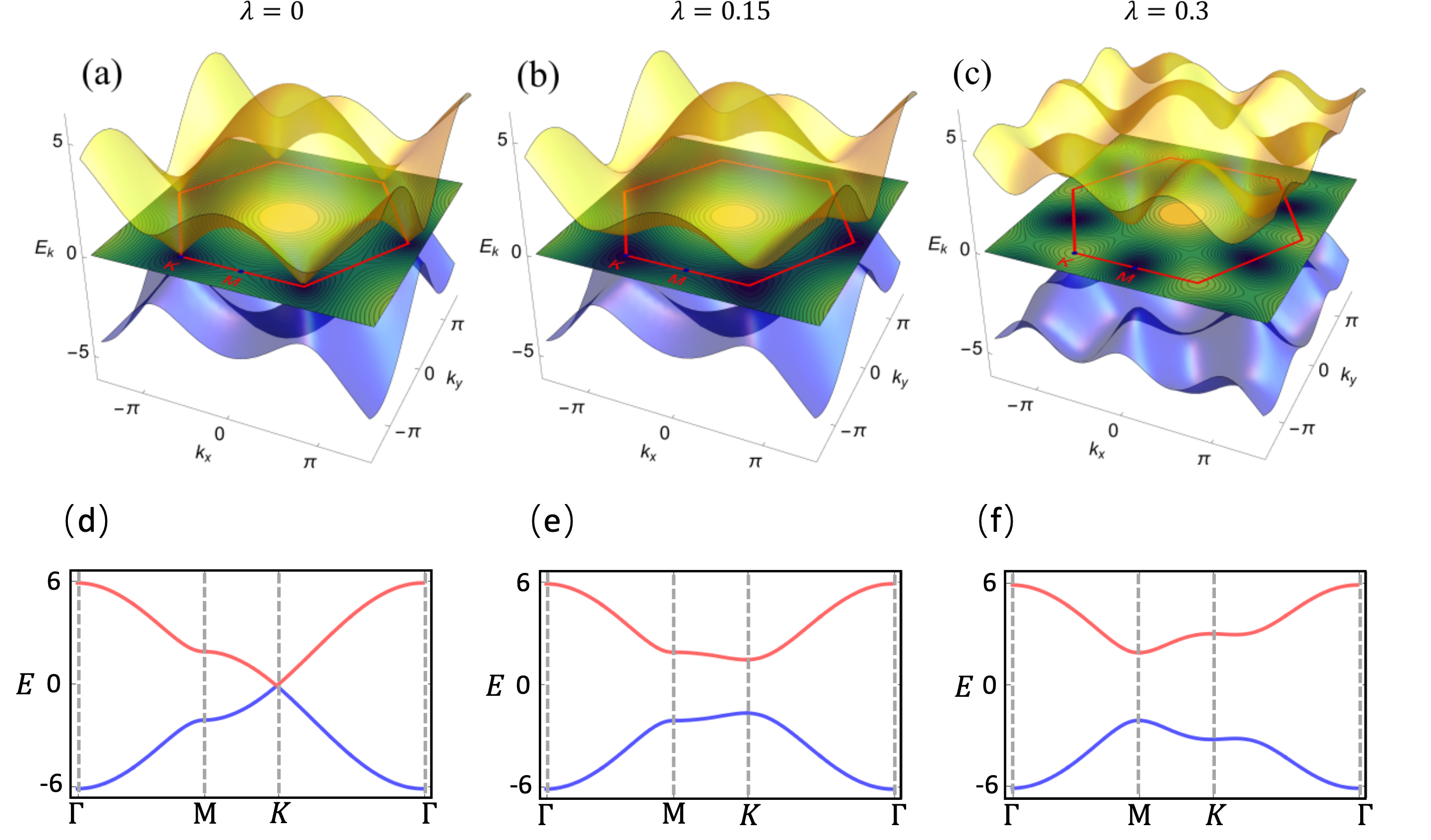}
    \caption{BdG band structure of free fermions and the cut along $\Gamma \rm M K \Gamma$ in the zero-flux sector. The TR-breaking perturbation are respectively (a,d) $\lambda = 0$, i.e. the pure KSL, (b,e) $\lambda=0.15$, (c) $\lambda=0.3$. The low energy soft mode remains at the $\rm M$ point with $\Delta E_{k=\rm M} = 2$, as the modes in $\rm K$ points are pushed to higher energy at larger $g$. The inset 2D planes in (a-c) are contours of the upper band, for better visualization of the locations of low-energy modes. Note that under small perturbation $\lambda$ in (b,e), the lowest-lying modes locate at $\pm \rm K$; while for larger perturbation in (c,f), the lowest-energy modes would locate at $\rm M$.  }
    \label{fig:bands}
\end{figure}
%%%%%%%%%%%%%%%%%%%%%%%%%%%%%%%%%%%%%%%%%%%%

We also present the varying band structures in this formalism for various TR-breaking perturbations in Supplementary Figure~\ref{fig:bands}. Importantly, for a small perturbation $\lambda$, the lowest-lying modes are located at the $\pm \rm K$ point, while for larger strengths of the perturbation, the lowest-energy modes move to the $\rm M$ points as shown in Supplementary Figure~\ref{fig:bands}(c,d). This is because the perturbation term evaluates to zero at $\rm M$, lifting the energy of modes at $\rm \pm K$ above that at $\rm M$ which is left unaffected. This explains the gapless modes at $\rm M$ immediately after the transition into the IGP, because under the large perturbation it is the soft mode at $\rm M$ with the minimal gap that is the most susceptible to the presence of fluxes.  We will revisit this in Supplementary Figure~\ref{fig:specfunc} of \ref{sec:spec}.

\section{Validity of mean field theories}
To understand the essence of the emergent IGP, it is imperative to introduce a proper MFT which moves us from an expansive tracing of a many-body tensor network (iPEPS) state to a practical quasi-particle mean-field picture. For this purpose, it is crucial to identify the primary degrees of freedom that govern the emergent intermediate phase. In Ref.~\cite{wang2024} we have highlighted the pivotal role of strong fluctuations in the flux sector for the IGP's emergence. This insight suggests a mean-field construct where fluxes and Majorana fermions are considered the elementary constituents. It is pertinent to note that this proposed mean field ansatz is distinct from existing MFTs aimed at understanding the intermediate phase \cite{NasuPRB2018,LiangPRB2018,Jiang2020,ZhangNatComm2022}. While these existing theories qualitatively address the occurrence of the field-induced quantum phase transitions of the Kitaev honeycomb model, they fall short in accurately capturing the essence of the emergent quantum phases in intermediate magnetic fields. Their limitation lies in an inability to describe the critical many-body entanglement amongst fractionalized degrees of freedom. 

In Kitaev's decomposition, the many-body Hamiltonian is expressed by four types of Majorana fermions $H = \sum_{\expval{ij},\alpha} ib_i^\alpha b_j^\alpha c_i c_j$, where $ib_i^\alpha b_j^\alpha \equiv i \hat{u}_{ij}^\alpha$ become the $\mathbb{Z}_2$ gauge fields in the extended Hilbert space at the pure Kitaev limit. When subjected to an external magnetic field, $ib_i^\alpha b_j^\alpha$ no longer commutes with the Hamiltonian, thus fluxes are no longer conserved. To study the quantum phase transition under a non-perturbative field, one may, for example, solve self-consistently equations whereby $\expval*{b_i^\alpha b_j^\alpha}$ are chosen to be mean fields, as was done in Ref.~\cite{ZhangNatComm2022,Jiang2020}.  The deficiency of such MFT ansatz is that it presumes that $b_i^\alpha b_j^\alpha$ for $\alpha = x,y,z$ can be approximated by product states devoid of higher-order entanglement between different flavors of fermions, which necessarily requires a four-fermion constraint that cannot be applied exactly. Now the flux excitation is characterized by an intrinsic six-body entanglement that cannot be reduced to fewer-body counterparts \cite{levin2006detecting,Preskill2006,feng2023stat}. For example, the presence of a vison requires $W_p = -1$,  this necessarily requires a sixth-order irreducible correlation characterized by
\begin{equation}
    W_p = u_{21} u_{23} u_{43} u_{45} u_{65} u_{61} = -1
\end{equation}
whereby there are $2^{6-1}$ possible $\mathbb{Z}_2$-field configurations on the plaquette. The $-1$ in the exponent is due to a six-body entanglement or the gauge invariance condition, that any one of the six $u_{ij}$'s, thus $\expval*{b_i^\alpha b_j^\alpha}$, must be conditioned on the other five. Hence, for any such parton decomposition to be faithful, it is necessary to either enforce the four Majorana constraint $ib_j^x b_j^y b_j^z c_j = 1~\forall j$, or apply the fermion parity projection that effectively serves as a many-body entangler \cite{Daniel2011,Knolle2019}. But these methods are, as is also mentioned in \cite{Jiang2020}, either computationally unfeasible or valid only in zero field limit. This explains why the existing MFTs based on the aforementioned parton decomposition have found conflicting results regarding the nature of the emergent intermediate phase, and are not able to capture its gapless nature as evidenced by iPEPS that is not affected by finite size limitations \cite{wang2024}. 

However, the above discussion does not necessarily imply that a MFT cannot be constructed, as we discuss below.
We demonstrate a properly constructed MFT that is capable of reproducing both the qualitative and quantitative characteristics of the IGP as revealed by iPEPS. The critical factor is the significant role played by a finite flux density for the emergence of IGP, as indicated in Ref.~\onlinecite{wang2024}. Given that the matter sector of CSL before $h_{c1}$ is equivalent to a gapped $p+ip$ superconductor, the finite-flux density in IGP and the concomitant gaplessness is equivalent to anyons with interactions of random signs, which is characterized by the mean field picture of a thermal Majorana metal \cite{Senthil2000,Chalker2001,Huse2012,Knolle2019}. Here, the term ``thermal" refers to an ensemble of flux configurations conditioned on a finite flux density, which effectively closes the Majorana gap through disorder average \cite{Huse2012}. These fluctuations are indeed of quantum origin rather than of thermal origin. Indeed, this picture is also supported by a recent DMRG study of the intermediate phase \cite{Baskaran2023}, where an apparent glassy phase of visons with a large flux density is found at intermediate fields. Furthermore, these visons are reported to exhibit slow dynamics such that the transport properties are primarily determined by (renormalized) Majorana fermions.  We can therefore write down the ansatz for IGP's ground state as
\begin{equation} \label{eq:mftapp}
    \ket{\Psi_{\rm IGP}} \simeq \sum_{\{\mathcal{F}\}} \psi_\mathcal{F} \ket{\mathcal{F}} \otimes \ket{\mathcal{M}_\mathcal{F}}
\end{equation}
where $\ket{\mathcal{F}}$ denote configurations in the flux sector in keeping with a finite flux average $\abs{\overline{W}_{p}} \ll 1$ in IGP, as implied in the ground state expectation of $\hat{W}_p$ at intermediate fields; and $\ket{\mathcal{M}_\mathcal{F}}$ the Majorana sector conditioned on the flux configuration. Both $\ket{\mathcal{F}}$ and $\ket{\mathcal{M}_\mathcal{F}}$ should be perceived as renormalized by mean fields. 
It is equivalent to a Schmidt decomposition of the many-body ground state which effectively separates the flux and Majorana degrees of freedoms. 
This ansatz, while phenomenological, does not suffer from the deficiencies of previous microscopic MFTs using parton decomposition, as Supplementary Equation~\eqref{eq:mftapp} preserves the high-order entanglement in $\ket{\mathcal{F}}$ whereby fluxes remain well-defined physical degrees of freedom. Hence the density matrix in the Majorana sector which governs the dynamical signatures of IGP can be written as an effective weighted ensemble:
\begin{equation}
    \rho_\mathcal{M} = \Tr_{\mathcal{F}} \ket{\Psi_{\rm IGP}}\bra{\Psi_{\rm IGP}} = \sum_{\{\mathcal{F}\}}\abs{\psi_\mathcal{F}}^2 \ket{\mathcal{M}_\mathcal{F}}\bra{\mathcal{M}_\mathcal{F}}
\end{equation}
where the weighted sum over $\abs{\psi_\mathcal{F}}^2$ characterizes a disorder averaging of Majorana sectors conditioned on the random 
ensembles of flux configurations. In the main text, we have elaborated on how the quantitative gapless signatures in energy and momentum of IGP emerge via the mechanism of the $p$-wave superconductor to metal transition.

\section{Majorana-hopping model with sign-disorder}
In our numerical calculations, we define $p_{\text{bond}}$ to be the probability of flipping $u_{ij}$ on each bond \textit{independently}; this allows us to generate random flux configurations. The ensemble average of the plaquette operator around a hexagon (i.e., $\overline{W}_{p}$) decreases monotonically with $p_{\text{bond}}$ as shown in Supplementary Figure~\ref{fig:SMF1}(a). For an ensemble of flux configuration $\{\ket{\mathcal{F}}\}$ generated by a given $p_{\text{bond}}$, the distribution of $W_{p}^{\mathcal{F}}$ is a Gaussian-like distribution as shown in Supplementary Figure~\ref{fig:SMF1}(b). This is because that $W_{p}^{\mathcal{F}}$ is essentially a summation of $\overline{W_{p}}$ values within all hexagons, which can be viewed as independent identical variables.According to the central limit theorem, this ensures that the distribution of $W_{p}^{\mathcal{F}}$ must be Gaussian.

While we have utilized a specific method to generate random flux configurations in which $W_{p}^{\mathcal{F}}$ follows a Gaussian distribution, we emphasize that the emergence of the Majorana metal phase from the chiral spin liquid (CSL) does not depend on the specific details of these random flux configurations. This robustness stems from the fact that each flux traps a Majorana fermion, and that flux proliferation results in a gapless Majorana metal phase, independently of the details of the distribution of $W_{p}^{\mathcal{F}}$. To support this, we generate an alternative set of random flux configurations in which $W_{p}^{\mathcal{F}}$ is fixed (rather than Gaussian-distributed) across all samples in the ensemble. As shown in Supplementary Figure~\ref{fig:SMF1}(c)-(e), the Majorana metal phase still emerges at low values of $\overline{W}{p}=W{p}^{\mathcal{F}}$. 

\begin{figure}[h]
    \centering
\includegraphics[width=0.9\linewidth]{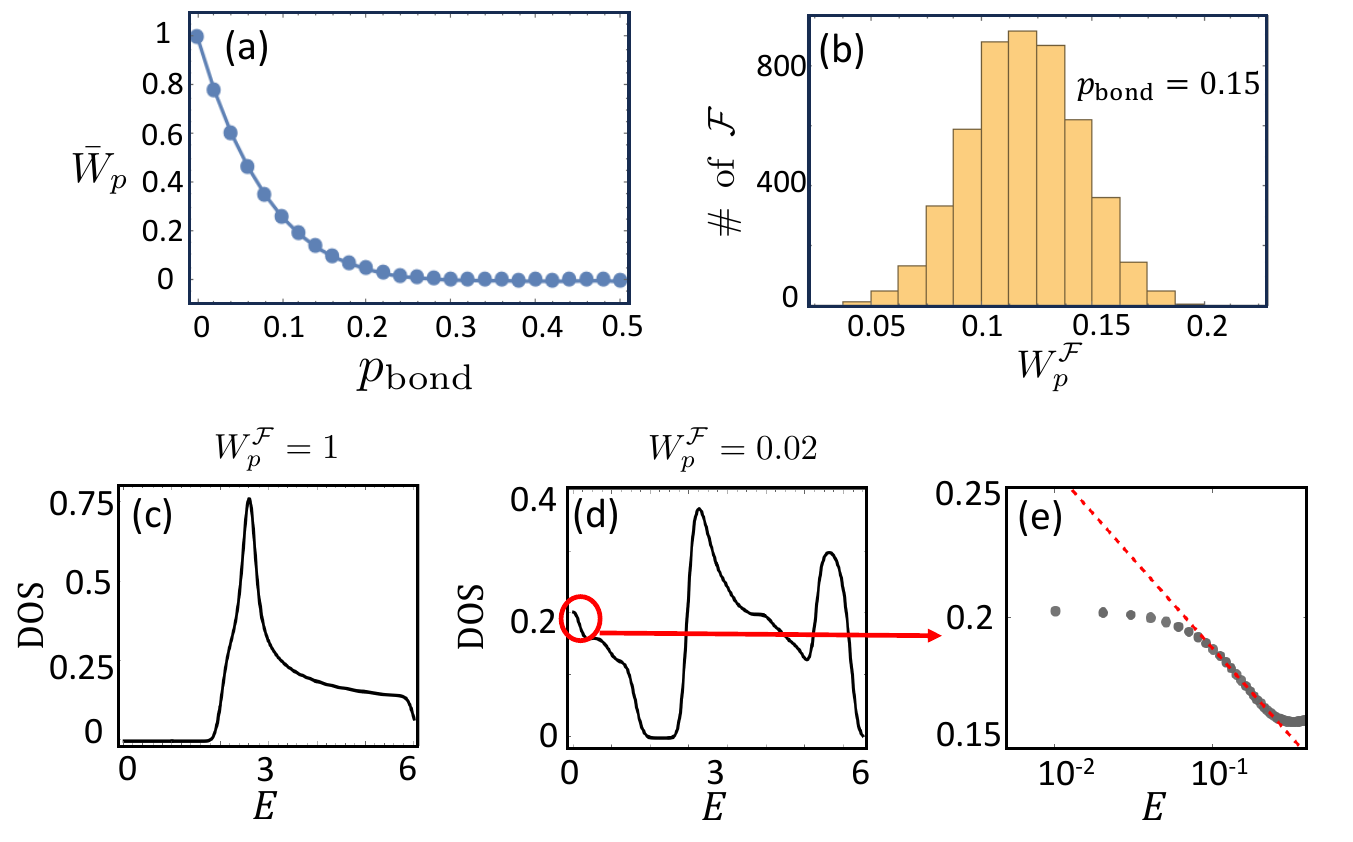}
    \caption{(a)The plot of $\overline{W}_{p}$ versus $p_{\text{bond}}$. (b) The distribution of $W^{\mathcal{F}}_{p}$ for a random ensemble of flux configurations generated by $p_{\text{bond}}=0.15$ with 5000 samples.(c)DOS when there is no flux disorder. (d) Averaged DOS for an ensemble of flux configuration with fixed $W_{p}^{\mathcal{F}}=0.02$. (e) The logarithm scaling of DOS at low energy.}
    \label{fig:SMF1}
\end{figure}

\section{The proliferation of flux-trapped Majorana zero modes}
\subsection{$\pi$-fluxed trapped Majorana zero modes}
In the main text, we use the fact that the Majorana sector of the CSL phase can be interpreted as a p-wave topological superconductor to argue that there should be Majorana zero modes (MZMs) trapped at the $\pi$-flux\cite{leezhangxiang07}. 
Here, we present an alternative demonstration along with numerical calculations to support this assertion.

\begin{figure}[h]
    \centering
\includegraphics[width=0.7\linewidth]{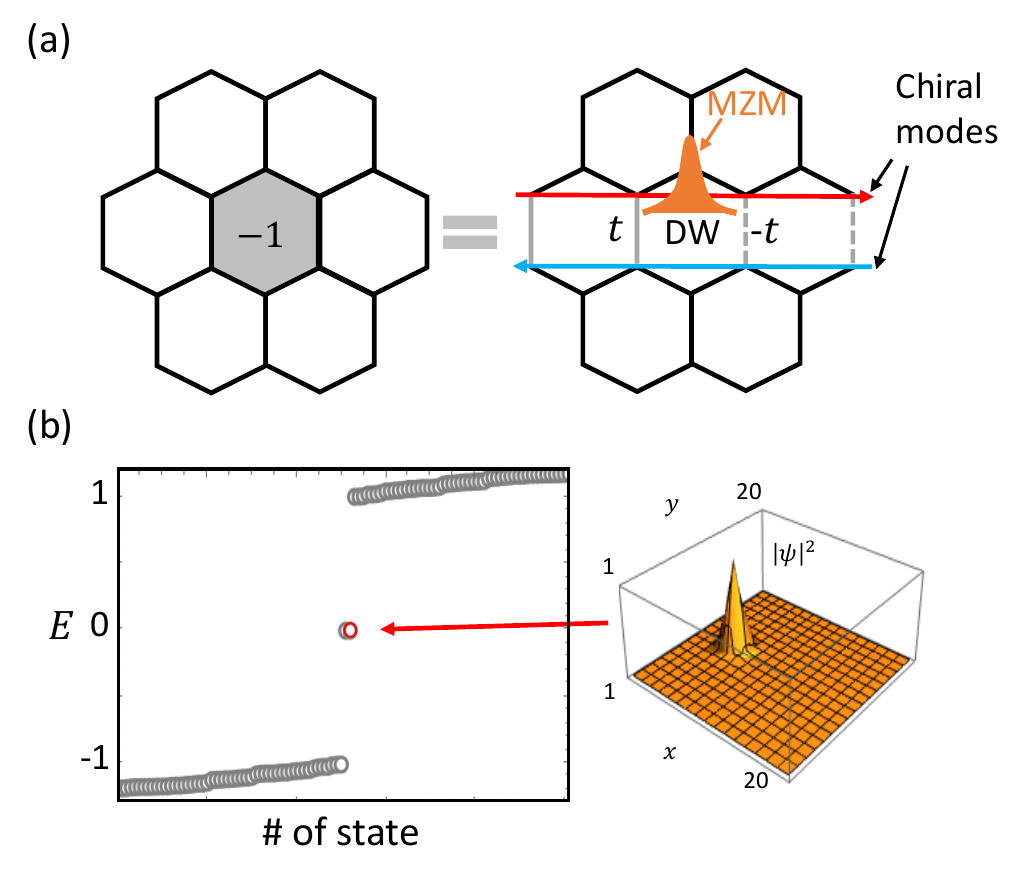}
    \caption{(a) Illustration the presence of a domain wall at a $\pi$-flux. (b) Energy spectrum of a system with a pair of $\pi$-fluxes far away from each other.  The wavefunction profile of one MZM is shown on the side}
    \label{fig:SMF2}
\end{figure}

We start our demonstration by identifying that inserting a single flux in the model in Eq.~(1) in the main text is equivalent to connecting two patches of the honeycomb lattice with a position depended hopping $t(x)$, as illustrated in Supplementary Figure~\ref{fig:SMF2}(a). If we turn off $t(x)$ between the two patches,  there will be gapless chiral Majorana modes on the boundary of both patches according to the bulk-boundary correspondence associated with Chern bands. The two boundary chiral Majorana modes together can be effectively described by a 1D Dirac Hamiltonian at low energy: $H_{0\text{eff}}=k_{x}\sigma_{z}=-i\partial_{x}\sigma_{z}$. Upon turning on $t(x)$, the two boundary chiral Majorana modes get coupled together and acquire a mass term, $t(x) \sigma_{x}$ in the Dirac Hamiltonian, leading to 
\begin{equation}
\label{eq:HDirac}
H_{\text{eff}}=-i\partial_{x}\sigma_{z}+t(x) \sigma_{x},
\end{equation}
where  $t(x)$ changes sign and creates a domain wall at the flux ($x=x_0$). Note that the coupling between the two patches does not introduce mass terms proportional to $\sigma_{y}$ because of particle-hole symmetry generated by the complex conjugate operator $K$. Next, we show from Supplementary Equation~\eqref{eq:HDirac} that a MZM is trapped at the domain wall described by an ansatz $\psi=e^{-\alpha(x-x_{0})}u$
which has a zero eigenvalue, \begin{equation}
H\psi=(i\alpha \sigma_{z}+t(x)\sigma_{x})u=0.
\end{equation}
Multiplying the above equation by $\sigma_{x}$ on both side leads to
\begin{equation}
\alpha \sigma_{y}u=-t(x)u,
\end{equation}
of which the localized solution corresponds to $u=(1 \ i)^{\rm T}/\sqrt{2}$, and $\alpha=-t(x)$ that is positive (negative) when $x>x_{0}$ ($x<x_{0}$). The localization length of this MZM is determined by the bulk gap since both $|t(x)|$ and the localization length of the chiral edge modes are directly related to the bulk gap. 

In Supplementary Figure~\ref{fig:SMF2} (b), we plot the energy spectrum of a periodic system containing two $\pi$-fluxes that are widely separated. Indeed, two MZMs have been observed, and the wavefunction profile of one of them explicitly manifests its localized nature.

\subsection{Flux Proliferation}
In the main text, we propose that when the average separation between two $\pi$-fluxes (i.e., 1/ $\sqrt{\text{flux density}}$) is comparable to the localization length of the Majorana zero modes (MZMs), the MZMs will strongly couple with each other and form a gapless band. Here, we provide numerical evidence supporting this picture. In Supplementary Figure~\ref{fig:SMF3}(a), we plot the density of states (DOS) at zero energy as a function of $1-\overline{W}_{p}$ for various values of the next-nearest hopping parameter $\lambda$. Here, $1-\overline{W}_{p}$ is positively correlated with the external magnetic field strength, while $\lambda$ determines the bulk gap of the Majorana hopping model at the $K$ point for $\lambda < 0.2$. Systems with larger $\lambda$ exhibit slower crossovers with external magnetic field. To see this more clearly, we plot $1-\overline{W}_{p,0}$ versus $\lambda$ in Supplementary Figure~\ref{fig:SMF3} (b), where $\overline{W}_{p,0}$ is the ensemble-averaged flux that makes DOS at zero energy half of the saturation value. Indeed we observe that $1-\overline{W}_{p,0}$ first increases with $\lambda$, then saturates around $\lambda=0.2$, after which $\lambda$ no longer controls the bulk gap.
\begin{figure}[h]
    \centering
\includegraphics[width=0.7\linewidth]{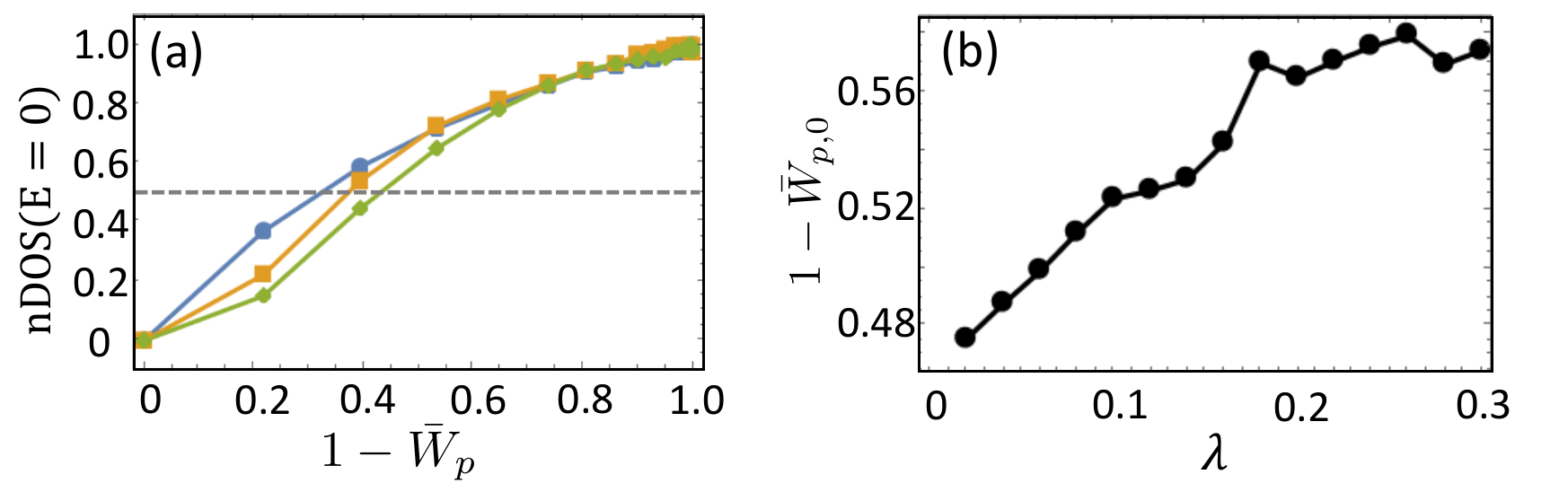}
    \caption{(a) Increase of DOS at zero energy with proliferation of $\pi$-flux. (b) Plots of $1-\overline{W}_{p,0}$ where DOS at zero energy is half of its saturation value for different $\lambda$s. nDOS means the normalized DOS at zero energy, and the normalization is done by dividing the saturation value. Each data point is derived by averaging over 100 samples.}
    \label{fig:SMF3}
\end{figure}

\begin{figure}[h]
    \centering
\includegraphics[width=1\linewidth]{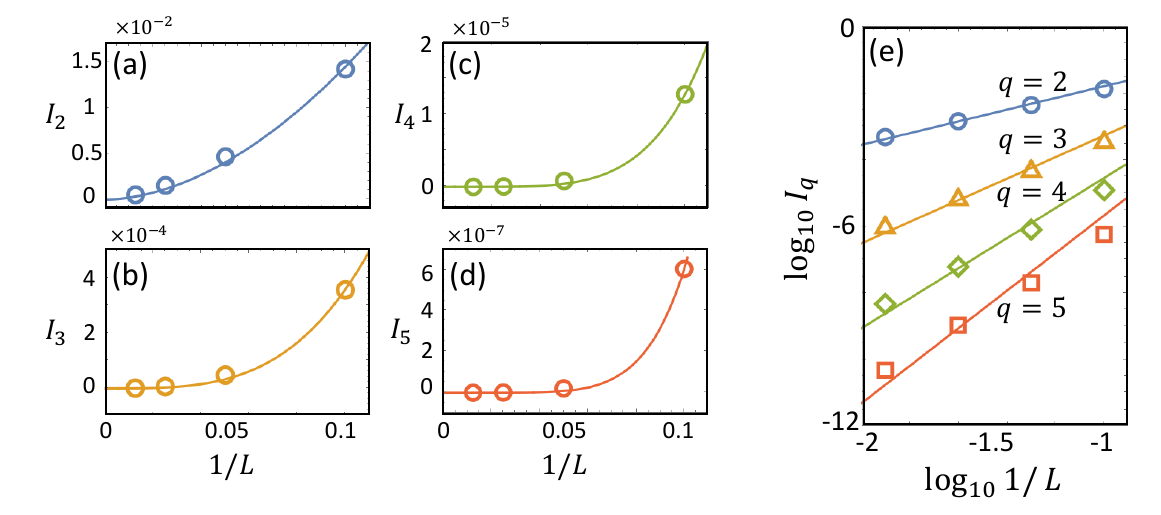}
    \caption{(a)-(d) show $I_{q=2,3,4,5}$ plotted against $1/L$. (e) show a logarithmic plot of $I_{q}$ and $1/L$, from which we fit the $\tau_{q}$ to be $1.76, 3.24, 4.52, 5.65$ for $q=2,3,4,5$. $\lambda=0.25$ and $\overline{W}_{p}=0.05$ are used for the calculations.}
    \label{fig:IPR}
\end{figure}

\section{Inverse participation ratio of the zero energy states in Majorana metal}
Random Majorana-hopping models of class D are known to exhibit three distinct phases: a topological insulator, a trivial insulator, and a gapless metal phase dubbed as Majorana metal. The topological insulator phase corresponds to the CSL phase of the Kitaev honeycomb model, where the band of Majorana fermions possesses a nonzero Chern number. In this work, we have identified the IGP in the Kitaev honeycomb model under a moderate magnetic field as the Majorana metal phase. 
Previous renormalization group analyses \cite{Senthil2000} have established that the Majorana metal phase hosts extended states at zero energy (the Fermi energy), with a density of states scaling logarithmically with energy, i.e., as $\ln |E|$. The disorder averaged DOS versus $E$ directly manifests the $\ln |E|$ scaling, verifying the phase under investigation is indeed a Majorana metal phase. 

To see explicitly that the state in IGP corresponds to the metallic (extended) phase instead of localized phase, we further calculate the inverse participation ratio (IPR) $I_{q}$ with $q=2,3,4,5$ of states at zero energy, which explicitly reveals the extended and multifractal nature of zero energy states. $I_{q}$ is defined as
\begin{equation}
\label{eq:IPR}
I_{q}=\int d\mathbf{r}|\psi(\mathbf{r})|^{2q},
\end{equation}
for a normalized wavefunction $\psi(\mathbf{r})$. In our calculations for finite-size systems, we first compute the averaged IPR for eigenstates in a small energy window around zero energy, and then do sample average to generate the plot in Supplementary Figure~\ref{fig:IPR}. We observe that (i) $I_{q}$ goes to zero as the linear size $L$ of the system goes to infinity [c.f. Supplementary Figure~\ref{fig:IPR}(a)-(d)]; and (ii) $I_{q}\sim 1/L^{\tau_{q}}$ with $\tau_{q}$ to be different for different $q$'s [c.f. Supplementary Figure~\ref{fig:IPR}(e)]. In this sense, the eigenstates at zero energy are extended and multifractal, which serves as direct evidence for the presence of the Majorana metal phase. Our result is consistent with previous investigation into disorder-induced Majorana metal \cite{Huse2012}.

\section{Justifications of Eq.~(5): Relating the structure factor to Majorana spectra}
\label{sec:eq5}
In this section we justify Eq.~(5) in the main text, and show how the spin-spin correlations are related to the Majorana spectral function. 

Upon substituting our ansatz $\ket{\Psi_{\rm IGP}}$ into Eq.~(5) in the main text, leads to
\begin{equation}
\begin{aligned}
\label{eq:spincorrproof}
    \expval*{\boldsymbol{\sigma}_i(t)\cdot \boldsymbol{\sigma}_j}&=\sum_{\mathcal{F},\mathcal{F}^{\prime},\alpha=x,y,z}\psi^{\star}_{\mathcal{F}^{\prime}}\psi_{\mathcal{F}}e^{iE_{0}t}\bra{\mathcal{F}^{\prime}}\otimes\bra{\mathcal{M}_{\mathcal{F}^{\prime}}}\sigma^{\alpha}_{i}e^{-i Ht}\sigma^{\alpha}_{j}\ket{\mathcal{F}}\otimes\ket{\mathcal{M}_{\mathcal{F}}}
    \\  &=\sum_{\mathcal{F},\mathcal{F}^{\prime},\alpha=x,y,z}\psi^{\star}_{\mathcal{F}^{\prime}}\psi_{\mathcal{F}}e^{iE_{0}t}\bra{\mathcal{F}^{\prime}_{\alpha;i}}\otimes\bra{\mathcal{M}_{\mathcal{F}^{\prime}}}c_{i}e^{-i Ht}c_{j}\ket{\mathcal{F}_{\alpha;j}}\otimes\ket{\mathcal{M}_{\mathcal{F}}}
    \\
    &=\sum_{\mathcal{F},\mathcal{F}^{\prime},\alpha=x,y,z}\psi^{\star}_{\mathcal{F}^{\prime}}\psi_{\mathcal{F}}e^{iE_{0}t-iE_{0\mathcal{F}}t}\delta_{\mathcal{F}^{\prime}_{\alpha;i}\mathcal{F}^{\phantom{\prime}}_{\alpha;j}}\bra{\mathcal{M}_{\mathcal{F}^{\prime}}}c_{i}e^{-i H_{\mathcal{M}}t}c_{j}\ket{\mathcal{M}_{\mathcal{F}}}
    \\
    &=\sum_{\mathcal{F},\alpha=x,y,z}\psi^{\star}_{\mathcal{F}_{\alpha;ij}}\psi_{\mathcal{F}}\bra{\mathcal{M}_{\mathcal{F}_{\alpha;ij}}}c_{i}(t)c_{j}\ket{\mathcal{M}_{\mathcal{F}}}
\end{aligned}
\end{equation} 
where $\mathcal{F}_{\alpha;ij\ldots}$ represents the flux configuration derived by flipping the $\alpha$-bond at $i,j,\ldots$ in  $\mathcal{F}$. In the third step we have used $\bra{\mathcal{F}^{\prime}}e^{-iHt}\ket{\mathcal{F}}=e^{-iE_{0\mathcal{F}}t}\delta_{\mathcal{F}\mathcal{F}^{\prime}}e^{-i H_{\mathcal{M}}t}$. This equation is valid because the flux sector is glassy and has slow dynamics, i.e., $e^{-iHt}\ket{\mathcal{F}}$ is almost the same with $\ket{\mathcal{F}}$ for finite $t$. Next, we argue that $H_{\mathcal{M}}$'s for two flux configures that are different only locally at several hexagons are indistinguishable in the thermodynamic limit, because they differ from each other up to several local hopping terms that will be negligible in the thermodynamic limit. This indicates that $\langle\mathcal{M}_{\mathcal{F}_{\alpha;ij}}|\mathcal{M}_{\mathcal{F}}\rangle\sim 1$. We can further approximate $\psi^{\star}_{\mathcal{F}_{\alpha;ij}}\psi_{\mathcal{F}}$ by $|\psi_{\mathcal{F}}|^2$, and then Supplementary Equation~\eqref{eq:spincorrproof} becomes the flux-configuration average of $\bra{\mathcal{M}_{\mathcal{F}}}c_{i}(t)c_{j}\ket{\mathcal{M}_{\mathcal{F}}}$.

In the following, we show that the Fourier transformation of $\bra{\mathcal{M}_{\mathcal{F}}}c_{i}(t)c_{j}\ket{\mathcal{M}_{\mathcal{F}}}$ is the spectral function. We start by inserting a complete set of states
$\sum_{m}\ket{m}\bra{m}$:
\begin{equation}
\label{eq:corrtospec}
\bra{\mathcal{M}_{\mathcal{F}}}c_{i}(t)c_{j}\ket{\mathcal{M}_{\mathcal{F}}}=\sum_{m}e^{-iE_{m}t}\bra{\mathcal{M}_{\mathcal{F}}}c_{i}\ket{m}\bra{m}c_{j}\ket{\mathcal{M}_{\mathcal{F}}},
\end{equation}
where $E_{m}$ is the energy difference between $\ket{m}$ and the ground state. Note that the excitation states of $H_{\mathcal{M}}$ are Bogoliubov states $\gamma_{n}^{\dag}=\varphi_{in}c_{i}$. Then, $c_{i}$ can be decomposed in the Bogoliubov states basis: $c_{i}=\varphi^{\star}_{ni}\gamma_{n}^{\dag}$. Substituting this into Supplementary Equation~\eqref{eq:corrtospec}, we find that only $\ket{m}=\gamma_{n}^{\dag}\ket{\mathcal{M}_{\mathcal{F}}}$ contributes, and upon Fourier transforming of $t$, we obtain
\begin{equation}
\label{eq:corrtospec1}
\int e^{i\omega t}dt \bra{\mathcal{M}_{\mathcal{F}}}c_{i}(t)c_{j}\ket{\mathcal{M}_{\mathcal{F}}}=\sum_{n}\varphi_{ni}\varphi_{nj}^{\star}\delta(\omega-E_{n}).
\end{equation}
If we sum all the contributions from the cases where $i,j$ belong to the same sublattice, i.e., $i=(\mathbf{r}_{1},\alpha)$ and $j=(\mathbf{r}_{2},\alpha)$ with $\alpha=A,B$ and $A,B$ labeling the sublattices, then Supplementary Equation~\eqref{eq:corrtospec1} becomes the spectral function in Eq.~(6) in the main text. 
%We can also generally calculate for $i=(\mathbf{r}_1,\alpha)$ and $j=(\mathbf{r}_2,\beta)$, which can be directly compared with the iPEPS calculations for $S^{\alpha}_{ab}(\mathbf{k},\omega)$, as shown in Supplementary Figure ~\ref{fig:}

\section{Spectral function} \label{sec:spec}
Here we show a detailed derivation of Eq.~(7) in the main text.

\begin{equation}
\label{eq:specfuncderive}
    \begin{split}
    A(\mathbf{r}_1,\mathbf{r}_2,\omega) = -\frac{1}{\pi} \mathfrak{Im}[G^+(\mathbf{r}_1,\mathbf{r}_2,\omega)]
        = \sum_n \operatorname{Tr}_{\text{cell}}\phi_n(\mathbf{r}_1) \phi_n^*(\mathbf{r}_2) \delta(\omega - E_n).
    \end{split}
\end{equation}
Note that we are considering a lattice with discrete positions $\mathbf{r}_{1}$ and $\mathbf{r}_{2}$. The trace over a unit cell implies that we consider the contributions of all internal degrees of freedom. As mentioned in \ref{sec:eq5}, this trace over unit cell corresponds to the sum of AA and BB spin-spin correlations. To explore the spectral function in momentum space, we first decompose each eigenstate $\phi_{n}(\mathbf{r})$ into a linear combination of Bloch states:
\begin{equation}
\label{eq:decomp2}
\phi_{n}(\mathbf{r})=\frac{1}{\sqrt{N}}\sum_{\alpha\mathbf{k}} c^{n}_{\alpha\mathbf{k}}e^{i\mathbf{k}\cdot \mathbf{r}}u_{\alpha\mathbf{k}},
\end{equation}
where $N$ is the number of unit cells in the system, and $\alpha$ labels sublattice indices. Then we substitute Supplementary Equation~\eqref{eq:decomp2} into Supplementary Equation~\eqref{eq:specfuncderive}, and rewrite the RHS of Supplementary Equation~\eqref{eq:specfuncderive} as
\begin{equation}
\label{eq:specfuncderive1}
    \begin{split}
\frac{1}{N}\sum_{n}\sum_{\alpha\alpha^{\prime}}\sum_{\mathbf{k}\mathbf{k}^{\prime}}c^{n}_{\alpha\mathbf{k}}e^{i\mathbf{k}\cdot\mathbf{r}_{1}}c^{n\star}_{\alpha^{\prime}\mathbf{k}^{\prime}}e^{-i\mathbf{k}^{\prime}\cdot\mathbf{r}_{2}}
\operatorname{Tr}_{\text{cell}}(u_{\alpha\mathbf{k}}u^{\dag}_{\alpha^{\prime}\mathbf{k}^{\prime}})\delta(\omega-E_{n}).
    \end{split}
\end{equation}
If we define $\mathbf{R}=(\mathbf{r}_{1}+\mathbf{r}_{2})/2$ and $\mathbf{r}=\mathbf{r}_{1}-\mathbf{r}_{2}$, then we can express the exponential terms in the above function as $e^{i(\mathbf{k}\cdot\mathbf{r}_{1}-\mathbf{k}^{\prime}\cdot\mathbf{r}_{2})}=e^{i(\mathbf{k}+\mathbf{k}^{\prime})\cdot\mathbf{r}/2+i(\mathbf{k}-\mathbf{k}^{\prime})\cdot\mathbf{R}}$, which is convenient for us to do the average over the center-of-mass position: 
\begin{equation}
\label{eq:specfuncderive2}
\begin{split}
\langle A(\mathbf{r}_1,\mathbf{r}_2,\omega)\rangle_{\mathbf{R}}  &\equiv \frac{1}{N}\sum_{\mathbf{R}}A(\mathbf{r},\mathbf{R},\omega)\\
&=\frac{1}{N^2}\sum_{\mathbf{R}}\sum_{n}\sum_{\alpha\alpha^{\prime}}\sum_{\mathbf{k}\mathbf{k}^{\prime}}e^{i(\mathbf{k}-\mathbf{k}^{\prime})\cdot\mathbf{R}}e^{i(\mathbf{k}+\mathbf{k}^{\prime})\cdot\mathbf{r}/2}c^{n}_{\alpha\mathbf{k}}c^{n\star}_{\alpha^{\prime}\mathbf{k}^{\prime}}\operatorname{Tr}_{\text{cell}}(u_{\alpha\mathbf{k}}u^{\dag}_{\alpha^{\prime}\mathbf{k}^{\prime}})\delta(\omega-E_{n})
\\
&=\frac{1}{N}\sum_{n}\sum_{\alpha\alpha^{\prime}}\sum_{\mathbf{k}\mathbf{k}^{\prime}}\delta_{\mathbf{k}\mathbf{k}^{\prime}}e^{i(\mathbf{k}+\mathbf{k}^{\prime})\cdot\mathbf{r}/2}c^{n}_{\alpha\mathbf{k}}c^{n\star}_{\alpha^{\prime}\mathbf{k}^{\prime}}\operatorname{Tr}_{\text{cell}}(u_{\alpha\mathbf{k}}u^{\dag}_{\alpha^{\prime}\mathbf{k}^{\prime}})\delta(\omega-E_{n})
\\
&=\frac{1}{N}\sum_{n}\sum_{\alpha\alpha^{\prime}}\sum_{\mathbf{k}}e^{i\mathbf{k}\cdot\mathbf{r}}c^{n}_{\alpha\mathbf{k}}c^{n\star}_{\alpha^{\prime}\mathbf{k}}\delta_{\alpha\alpha^{\prime}}\delta(\omega-E_{n})
\\
&=\frac{1}{N}\sum_{n\alpha\mathbf{k}}e^{i\mathbf{k}\cdot\mathbf{r}}|c^{n}_{\alpha\mathbf{k}}|^2\delta(\omega-E_{n})
\end{split}
\end{equation}
We have thus obtained Eq.~(7) in the main text. In Supplementary Figure~\ref{fig:specfunc}, we plot the sample-averaged (over 50 samples) spectral function along  high-symmetry directions for various $\overline{W}_{p}$ and $\lambda$. Fig.~2(c) in the main text corresponds to the low energy part of the lower right corner panel ($\lambda=0.25$ and $\overline{W}_{p}=0.05$) in  Supplementary Figure~\ref{fig:specfunc}, with spectral weight plotted on a logarithm scale. At the end of this part, we mention that besides taking trace over the internal degrees of freedom in the spectral function, we can alternatively sum over the anti-diagonal elements of the matrix corresponding to the internal degrees of freedom -- $u_{\alpha\mathbf{k}}u^{\dag}_{\alpha'\mathbf{k}}$. This will lead to a function corresponding to the sum of AB and BA spin-spin correlations, following the arguments in \ref{sec:eq5}. The sum of the spectral function and this new function can be directly compared with the total spin-spin correlations calculated in our previous work~\cite{wang2024spectral}, which are directly related to inelastic neutron scattering experiments.

%{\color{red}At the end of this part, we mention that besides taking trace over the internal degrees of freedom in the spectral function, we can alternatively sum over the anti-diagonal elements of the matrix corresponding to the internal degrees of freedom -- $u_{\alpha\mathbf{k}}u^{\dag}_{\alpha'\mathbf{k}}$. This will lead to a function corresponding to the sum of AB and BA spin-spin correlations, following the arguments in \ref{sec:eq5}. Then, as shown in Fig.~\ref{}, the sum of the spectral function and this new function can be directly compared with the total spin-spin correlations calculated in our previous work~\cite{wang2024spectral}, which are directly related to inelastic neutron scattering experiments. } 

\begin{figure}[h]
    \centering
\includegraphics[width=1\linewidth]{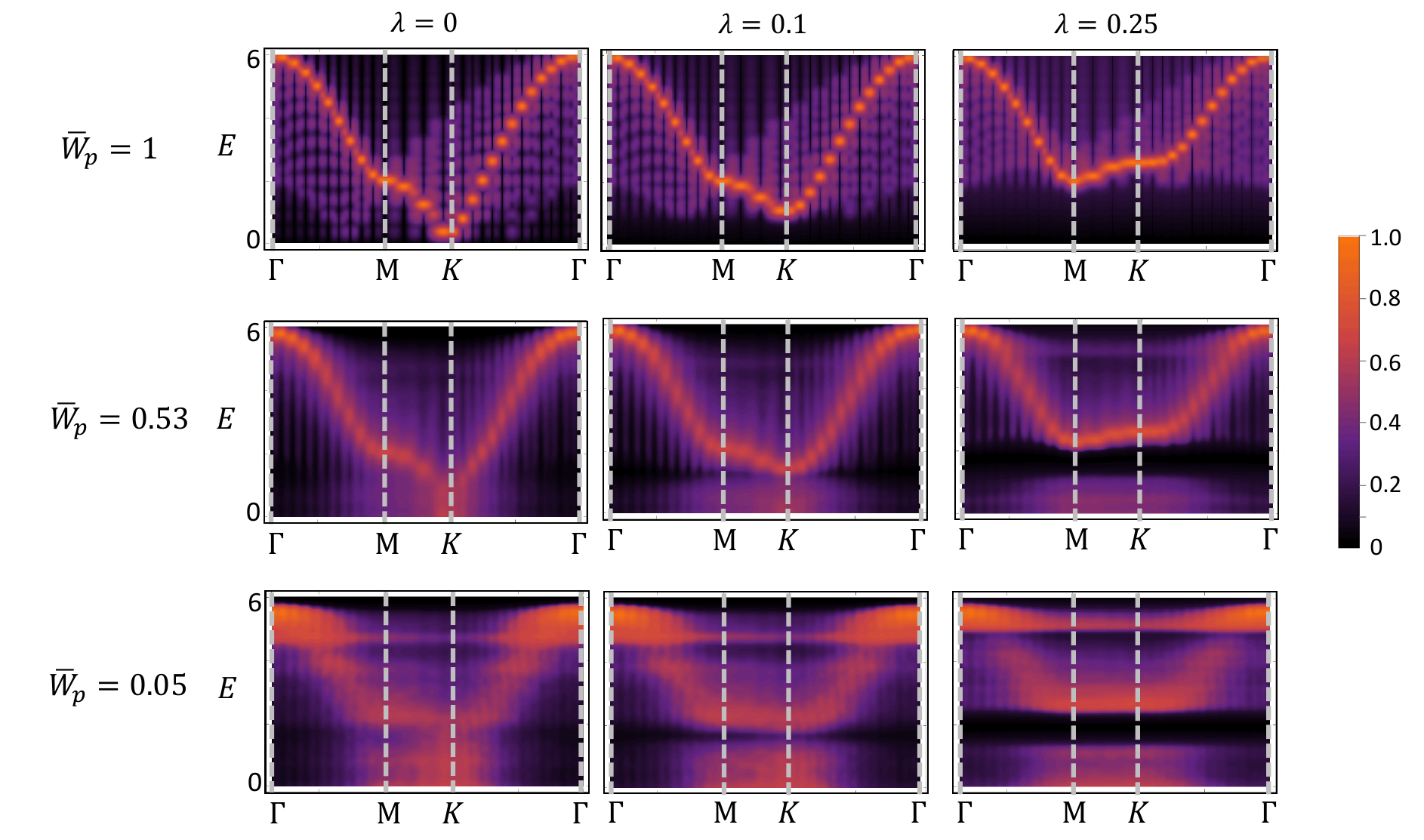}
    \caption{Energy and momentum resolved spectral function for various NNN hopping $\lambda$ and flux average $\overline{W}_{p}$ as an indicator of fluctuation (disorder) in the gauge field. under small perturbation $\lambda$, the lowest-lying modes locate at $\pm \rm K$; while for larger perturbation, the lowest-energy modes would locate at $\rm M$. At weak disorder, gapless modes first develop near $K$ for small NNN couplint ($\lambda = 0.1$); and near $M$ for larger NNN coupling ($\lambda=0.25$). As disorder gets stronger, gapless modes develop near both $\rm M$ and $\rm K$.}
    \label{fig:specfunc}
\end{figure}

\section{The four-Majorana spectrum}
Here we present the calculation for the correlation between four Majorana fermions under a single-mode approximation, relevant for Fig.~4 in the main text. The idea is to first obtain an exact formula for the Majorana band of the CSL phase, and modify the band structure such that the modes are gapless at momenta indicated by MFT and iPEPS calculation. This approach effectively models the emergent metallic bands induced by flux fluctuations, while stay in keeping with the Majorana content essential for the IGP. 
In the flux free sector we write the pure majorana Hamiltonian 
\begin{equation}
\label{eq:perturbedkitaev}
    H_0 = \sum_{j\in A,\delta} ic_{j,A} c_{j+\delta,B} - \sum_{j\in A, \delta} i \lambda\: c_{j,A} c_{j+\delta,A}
\end{equation}
where $\abs{\lambda} = h_x h_y h_z$. Here we have explicitly labelled the $A$ and $B$ sublattices for clearer definition of normal modes $\psi = (c_{k,A}, c_{k,B})$, and $\delta \in \{0, \mathbf{n}_1, \mathbf{n}_2\}$. In the $k$ space, we define two Majorana fields for each sublattice by $a_{\mathbf{k}} = \sum_j e^{i\mathbf{k} \cdot \mathbf{r}_j} c_{i,A}$ and $b_{\mathbf{k}} = \sum_j e^{i\mathbf{k}\cdot \mathbf{r}_j} c_{i,B}$, with $a_{\mathbf{k}}^\dagger = a_{-\mathbf{k}},~b_{\mathbf{k}}^\dagger = b_{-\mathbf{k}}$. 
The Hamiltonian becomes gapped upon introducing a NNN hopping as the leading order TR-breaking perturbation:
\begin{equation}
	H = \frac{1}{2}\sum_{\mathbf{q}}
	\begin{pmatrix}
		a_{-\mathbf{q}} & b_{-\mathbf{q}}
	\end{pmatrix}
	\begin{pmatrix}
		Q(\mathbf{q}) & if(\mathbf{q}) \\
		-i f^*(\mathbf{q}) & -Q(\mathbf{q})
	\end{pmatrix}
	\begin{pmatrix}
		a_{\mathbf{q}} \\ b_{\mathbf{q}}
	\end{pmatrix}
\end{equation}
where the TR is broken by \cite{kitaev2006anyons}
\begin{equation}
    Q(\mathbf{k}) = 4\lambda[\sin(\mathbf{k}\cdot \mathbf{n}_2) - \sin(\mathbf{k}\cdot \mathbf{n}_1) - \sin(\mathbf{k}\cdot \mathbf{n}_3)] 
\end{equation}
where $\mathbf{n}_3 = \mathbf{n}_2-\mathbf{n}_1$, $\lambda$ denotes the strength of the leading order TR-breaking perturbation. 
The off-diagonal elements for each mode are related to
\begin{equation}
    f(\mathbf{q}) =   J_x e^{iq_x} + J_y e^{iq_y} + J_z 
\end{equation}
where we've defined $q_x \equiv \mathbf{q}\cdot \mathbf{n}_1$ and $q_y \equiv \mathbf{q}\cdot \mathbf{n}_2$. We are interested in the isotropic Kitaev exchange, so from now on we will set $K_x = K_y = K_z = 1$. 
We investigate the dimer dynamics in the TR-breaking case. 
In the diagonal basis of the complex fermions $C$ we have \cite{Feng2022}
\begin{equation}
\begin{split} \label{eq:h0}
    H_{0} = \sum_{\mathbf{k}} E_\mathbf{k} \left( C_{\mathbf{k},1}^\dagger C_{\mathbf{k},1} - C_{\mathbf{k},2}^\dagger C_{\mathbf{k},2} \right) 
\end{split}
\end{equation}
The energy is given by $E_\mathbf{k} = \sqrt{Q_\mathbf{q}^2 + \abs{f(\mathbf{q})}^2}$, which is consistent with that in Supplementary Equation~\eqref{eq:Ek}; 
and the ground state is given by filling the band of $C_2$ fermion $\ket{\rm gs} = \prod_\mathbf{k} C_{\mathbf{k},2}^\dagger\ket{0}$. 
\begin{figure}[t]
    \centering
    \includegraphics[width=0.7\linewidth]{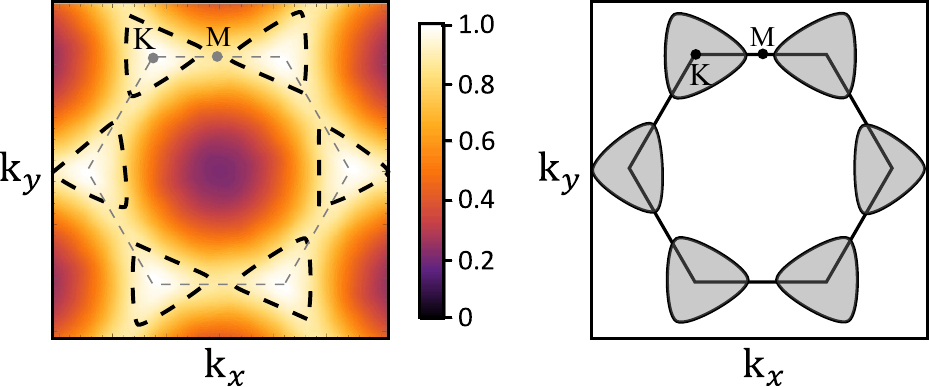}
    \caption{(Left) Mean-field results for the Majorana spectral function at zero energy $A(\mathbf{k}, \omega  =0)$ (i.e., the Majorana Fermi surface) in the IGP,  with $\pi$-flux density set by $\overline{W}_{p} = 0.01$. The gray dashed line represents the boundary of the first BZ, and the black dashed circles mark out the boundaries of regions that harbor the strongest signals near $\rm \pm K$ points. (Right) Our proposed gapless Majorana Fermi surface at $E_\mathbf{k} = 0$ in keeping with the MFT prediction shown on the left. The solid black line represents the boundary of the first BZ, and the gray blobs centered at $\pm \rm K$ points denote the presence of gapless Majorana modes around zero energy. }
    \label{fig:fourM}
\end{figure}
In the flux-free sector the two-Majorana operator $\mathcal{D}_j \equiv ic_j c_{j+z}$ can be written in momentum space:
\begin{equation} \label{eq:dk}
    \mathcal{D}_\mathbf{k} = {\rm F.T.}\{ic_j c_{j+z}\} = i\sum_\mathbf{q} a_{\mathbf{k} -\mathbf{q}} b_{\mathbf{q}}
\end{equation}
In the momentum-space and in terms of the complex fermion $C_{\mathbf{q},2}$ of Supplementary Equation~\eqref{eq:h0}:
\begin{equation}
\begin{split}
    \mathcal{D}_\mathbf{k}
    = i \sum_\mathbf{q}\frac{1}{2}\frac{\Delta_{\mathbf{k}-\mathbf{q}} - i\epsilon_{\mathbf{k}-\mathbf{q}}}{E_{\mathbf{k}-\mathbf{q}} - Q_{\mathbf{k}-\mathbf{q}}} \left(C_{{\mathbf{k}-\mathbf{q}},2} C_{\mathbf{q},2} + C_{-\mathbf{q},2}^\dagger C_{{-\mathbf{k}+\mathbf{q}},2}^\dagger \right)
\end{split}
\end{equation}
where $\epsilon_\mathbf{q}$ and $\Delta_\mathbf{q}$ are defined by
\begin{align}
	\epsilon_\mathbf{q} &= \mathfrak{R}[f(\mathbf{q})] =  2(\cos q_x + \cos q_y + 1),\\
	\Delta_\mathbf{q} &= \mathfrak{I}[f(\mathbf{q})] = 2(\sin q_x + \sin q_y)
\end{align}
Here we have ignored terms like $C_{\mathbf{k}+\mathbf{q}}C_{-\mathbf{q}}^\dagger$ which annihilates the ground state (as the filled Fermi sea), thus do not contribute to the dynamics. Hence, using Supplementary Equation~\eqref{eq:dk}, the dynamical four-Majorana correlation becomes \cite{wang2024}
\begin{equation} \label{eq:nnkmapp}
\begin{split}
        S_{2}(\mathbf{k},\omega)
        = \sum_m \mel{\rm gs}{\mathcal{D}_\mathbf{k}}{m} \mel{m}{\mathcal{D}_{-\mathbf{k}}}{\rm gs} \delta[\omega - (E_{\mathbf{k} - \mathbf{q}} + E_{-\mathbf{q}}) ]
        = \sum_{\mathbf{q} \in \rm BZ} W(\mathbf{k},\mathbf{q}) \delta[\omega - (E_{\mathbf{k} - \mathbf{q}} + E_{-\mathbf{q}}) ]
\end{split}
\end{equation}
where $\ket{m} \equiv \ket{m(\mathbf{k}, \mathbf{q})}$ are excited states associated with quasi-particles with momentum $\mathbf{k}$ and $\mathbf{q}$, and we have defined the four-Majorana spectral weight, as is used in the main text, $W(\mathbf{k},\mathbf{q})$ as 
\begin{equation} \label{eq:wkq}
    W(\mathbf{k},\mathbf{q}) = \frac{1}{4}\frac{E_{\mathbf{k}-\mathbf{q}}^2}{E_{\mathbf{k}-\mathbf{q}}^2 - Q_{\mathbf{k}-\mathbf{q}}^2} 
\end{equation}
See also Ref.~\cite{wang2024} and the supplemental material thereof. 
Equation \eqref{eq:nnkmapp} and Supplementary Equation~\eqref{eq:wkq} represent a generic formula for the four-Majorana spectrum of a Majorana band $E_\mathbf{k}$ on a honeycomb lattice, with NN an NNN Majorana hoppings under class D \cite{Senthil2000,Chalker2001,Huse2012}. Therefore, we can now use the input from the Majorana spectal function at zero energy $A(\mathbf{k}, \omega  =0)$ obtained with our MFT calculation for an approximated gapless band $E_\mathbf{k}$ which determines the four-Majorana spectra in Supplementary Equation~\eqref{eq:nnkmapp}. In the main text, we tuned the Majorana band $E_{\mathbf{k}}$ so as to give the desired zero-energy modes around $\pm \rm K$ points, as is predicted by our MFT ansatz. The $A(\mathbf{k}, \omega  =0)$ from MFT and our modeled Majorana band at $E_\mathbf{k} = 0$ is presented in Supplementary Figure~\ref{fig:fourM}. 

As an approximation, we treat the modes enclosed by the dotted loops in Supplementary Figure~\ref{fig:fourM} as zero energy modes. Specifically, we assume the Majorana band to be nearly flat and at zero energy near $\pm K$ points, and fix the energy of other modes in the first BZ based on the mini-gap information depicted in Fig.~2(c) of the main text. We then obtain the four-Majorana spectrum analytically under this single-mode approximation with a relatively large broadening factor $\sim 0.2$ to approximate the highly diffusive spectrum of Majorana fermions in energy. Our result plotted in Fig.~3(c) is in very good agreement with the unbiased numerical data obtained by iPEPS, as is shown in Fig.~4 of the main text, validating our claim that the IGP is an effective Majorana metal with a finite neutral Fermi surface.  Though in the single mode approximation we approximate all modes near $\pm K$ points to be at zero energy, we refrain from making definitive assertions regarding whether the FS precisely forms a loop with codimension $1$ or a disk with codimension $0$. However, it's crucial to note that in both cases, the Fermi surfaces can be robust, according to a topological classification of Fermi surfaces in class D \cite{topological2013zhao}.

\section{iPEPS Algorithm}
This section provides information about the iPEPS utilized for deriving the dynamical spectrum within the IGP. This method is the same as the algorithm used in our previous work \cite{wang2024}, which adopted iPEPS as the ansatz of the ground state and excited states. The ground state is described as infinite tensor network with translation invariant local tensor $A$. 

\begin{equation} 
    \ket{\psi_0} =\vcenter{\hbox{\includegraphics[scale=0.4]{PRL2/honeycomb.pdf}}}
\end{equation}

The Single-Mode Approximation (SMA), introduced in \cite{feynman1954atomic}, is employed to characterize excited states. SMA uses a variational ansatz for the excited state as shown below:
\begin{equation}
\ket{\Psi_\mathbf{k}}=\sum_{\mathbf{r}}e^{-i\mathbf{k}\cdot\mathbf{r}}\ket{\Psi_\mathbf{r}}=\sum_{\mathbf{r}}e^{-i\mathbf{k}\cdot\mathbf{r}}\vcenter{\hbox{\includegraphics[scale=0.4]{PRL2/exci.pdf}}}
\end{equation}
Here, $\mathbf{k}$ denotes momentum and $\ket{\Psi_\mathbf{r}}$ signifies the state with an excitation at site $\mathbf{r}$.
Under the representation of iPEPS,  it's implemented by replacing the local tensor $A$ of the ground state at site $\mathbf{r}$ with a disturbed local tensor $B$.

Excited states are required to adhere to the orthogonality constraint in relation to the ground state, depicted as $\braket{\psi_0}{\Psi_\mathbf{k}(B)}=0$. Additionally, they must eliminate gauge redundancy to ensure accuracy in calculations\cite{ponsioen2022automatic}.

After satisfying the two constraints, we can obtain the $B$ tensor by minimizing the cost function
\begin{equation}
    L = {\frac{\bra{\Psi_\mathbf{k}(B)}H-E_{\rm gs}\ket{\Psi_\mathbf{k}(B)}}{\braket{\Psi_\mathbf{k}(B)}{\Psi_\mathbf{k}(B)}}}.
\end{equation}
which can be interpreted as solving the equation:
\begin{equation}
    \frac{\partial}{\partial B}{\bra{\Psi_\mathbf{k}(B)}H-E_{\rm gs}\ket{\Psi_\mathbf{k}(B)}}=E_\mathbf{k}\frac{\partial}{\partial B}{\braket{\Psi_\mathbf{k}(B)}{\Psi_\mathbf{k}(B)}}
\label{min}
\end{equation}
where $E_\mathbf{k}$ is the excitation energy.

The dynamical spectral function at zero temperature is defined as follows:
\begin{equation}
\begin{aligned}
\label{spec}
    S^{\alpha\beta}(\mathbf{k},\omega)&=\bra{\psi_0}O^\alpha_{\mathbf{k}} \delta(\omega-H+E_{\rm gs}) O^\beta_{-\mathbf{k}}\ket{\psi_0}=\sum_m{\bra{\psi_0}O^\alpha_\mathbf{k}\ket{\Psi_\mathbf{k}^m}\bra{\Psi_\mathbf{k}^m}O^\beta_{-\mathbf{k}}\ket{\psi_0}}\delta(\omega-E_\mathbf{k}^m+E_{\rm gs}) 
\end{aligned}
\end{equation}
where $\alpha,\beta=x,y,z$. $\ket{\psi_m}$ denotes the eigenstate of hamiltonian $H$ with energy $E_m$. The corresponding spectral weight can be obtained by contracting the double-layer tensor
\begin{equation}
\begin{split}
\bra{\Psi_\mathbf{k}^m}O^\alpha_\mathbf{k}\ket{\psi_0}&=\frac{1}{\sqrt{N}}\sum_{\mathbf{rr'}}e^{-i\mathbf{k\cdot(r-r')}}\bra{\Psi_\mathbf{r'}^m}O^\alpha_\mathbf{r}\ket{\psi_0}.
\label{weight}
\end{split}
\end{equation}
The contraction can be obtained using the CTM summation introduced in \citeSM{ponsioen2022automatic}. The delta function in Supplementary Equation~\eqref{spec} can be approximated using the Lorentzian expansion with a broadening factor $\eta$. In all iPEPS calculations we fixed the bond dimension to $D=5$ and $\eta=0.05$ if not otherwise specified.

In the main text, we have focused on the spectrum of two kinds of excitations: one spin flip $\sigma_i^\alpha$ and two spin flip $\sigma_i^\alpha \sigma_{i+z}^\alpha (\alpha = x,y,z)$. We calculated the spectrum function in momentum space 
\begin{equation}
\begin{split}
S^{\alpha}_1(\mathbf{k},\omega)&=\frac{1}{N}\sum_{ij} e^{-i\mathbf{k
\cdot(R_i-R_j)}}[S_{aa}^{\alpha}(i,j,\omega)+S_{bb}^{\alpha}(i,j,\omega)]\\
S^{\alpha}_2(\mathbf{k},\omega)&=\frac{1}{N}\sum_{ij} e^{-i\mathbf{k\cdot(R_i-R_j)}}S_{2}^{\alpha}(i,j,\omega)
\end{split}
\end{equation}
where $\mathbf{R_i}$ and $\mathbf{R_i}$ represent the sites of unit cells, $a$ and $b$ are sublattice indices. Note that we consider only the intra-sublattice correlation in the two-spin correlation $S_{aa(bb)}$, which is in keeping with the spectral function obtained in our mean-field theory [Supplementary Equation\eqref{eq:specfuncderive2}] and is periodic within the first BZ. $S_{aa(bb)}^\alpha(i,j,\omega)$ and $S_{2}^\alpha(i,j,\omega)$ denote the spectral function in real space. In the Lehmann representation, we have:
\begin{equation}
\begin{split}
S_{aa(bb)}^\alpha(i,j,\omega)=\sum_m \bra{\psi_0}\hat{\sigma}_{i,a(b)}^\alpha\ket{\Psi_\mathbf{k}^m}\bra{\Psi_\mathbf{k}^m}\hat{\sigma}_{j,a(b)}^\alpha\ket{\psi_0}\delta(\omega-E_\mathbf{k}^m+E_{\rm gs})
\end{split}
\end{equation}
and
\begin{equation}
\begin{split}
S_{2}^\alpha(i,j,\omega)=\sum_m \bra{\psi_0}\hat{\sigma}_{i,a}^\alpha\hat{\sigma}_{i,b}^\alpha\ket{\Psi_\mathbf{k}^m}\bra{\Psi_\mathbf{k}^m}\hat{\sigma}_{j,a}^\alpha\hat{\sigma}_{j,b}^\alpha\ket{\psi_0}
         \delta(\omega-E_\mathbf{k}^m+E_{\rm gs}) 
\end{split}
\end{equation}
where we have excluded the ground-state contribution by setting $\hat{\sigma}_{i,a(b)}^\alpha=\sigma_{i,a(b)}^\alpha-\bra{\psi_0}\sigma_{i,a(b)}^\alpha\ket{\psi_0}$ and $\hat{\sigma}_{i,a}^\alpha\hat{ \sigma}_{i,b}^\alpha=\sigma_{i,a}^\alpha\sigma_{i,b}^\alpha-\bra{\psi_0}\sigma_{i,a}^\alpha\sigma_{i,b}^\alpha\ket{\psi_0}$.

% \section*{Supplementary References}
% \begingroup
% \renewcommand{\section}[2]{}%
%\renewcommand{\chapter}[2]{}% for other classes
% \bibliography{spectra}
\bibliographystyleSM{apsrev4-2}
\bibliographySM{spectra}  
% \endgroup

\end{widetext}

\end{document}